\documentclass[11pt]{article}
\usepackage{amssymb,amsmath,amsfonts}
\usepackage{graphicx}
\usepackage{graphics}
\usepackage{eepic,epsfig}

\@addtoreset{equation}{section}

\makeatother

\begin{document}

\title{Electromagnetic vacuum stresses and energy fluxes induced \\
by a cosmic string in de Sitter spacetime}
\author{A. A. Saharian$^{1}$\thanks{%
Corresponding author. E-mail address: saharian@ysu.am},\thinspace\ V. F.
Manukyan$^{1}$,\thinspace\ V. Kh. Kotanjyan$^{1,2}$, \thinspace\ A. A.
Grigoryan$^{1}$ \\
\\
\textit{$^1$Department of Physics, Yerevan State University,}\\
\textit{1 Alex Manoogian Street, 0025 Yerevan, Armenia}\vspace{0.3cm}\\
\textit{$^{2}$Institute of Applied Problems in Physics NAS RA,}\\
\textit{25 Hr. Nersessian Street, 0014 Yerevan, Armenia}}
\maketitle

\begin{abstract}
For the electromagnetic field in $(D+1)$-dimensional locally de Sitter (dS)
spacetime, we analyze the effects of a generalized cosmic string type defect
on the vacuum expectation value of the energy-momentum tensor. For the
Bunch-Davies vacuum state, the topological contributions are explicitly
extracted in both the diagonal and off-diagonal components. The latter
describes the presence of radially directed energy flux in the vacuum state.
It vanishes for $D=3$ because of the conformal invariance of the
electromagnetic field and is directed towards the cosmic string for $D\geq 4$%
. The topological contributions in the vacuum stresses are anisotropic and,
unlike to the geometry of a cosmic string in the Minkowski spacetime, for $%
D>3$ the stresses along the directions parallel to the string core differ
from the energy density. Depending on the planar angle deficit and the
distance from the cosmic string, the corresponding expectation values can be
either positive or negative. Near the cosmic string the effect of the
gravitational field on the diagonal components of the topological part is
week and the leading terms in the respective expansions coincide with the
expectation values for a cosmic string in background of Minkowski spacetime.
The spacetime curvature essentially modifies the behavior of the topological
terms at proper distances from the cosmic string larger than the dS
curvature radius. In that region, the topological contributions in the
diagonal components of the energy-momentum tensor decay in inverse
proportion to the fourth power of the proper distance and the energy flux
density behaves as inverse-fifth power for all values of the spatial
dimension $D$. The exception is the energy density in the special case $D=4$%
. For a cosmic string in the Minkowski bulk the energy flux is absent and
the diagonal components are proportional to the $(D+1)$th power of the
inverse distance.
\end{abstract}

\bigskip

\textit{Keywords:} Cosmic string, de Sitter spacetime, electromagnetic vacuum

\section{Introduction}

The canonical quantization of fields is based on the expansion of the field
operator in terms of a complete set of mode functions being the solutions of
the classical field equation. That expansion defines the annihilation and
creation operators and then the space of the Fock states is constructed. The
vacuum is defined as a state of a quantum field that is nullified by the
action of the annihilation operator. The mode functions employed in the
quantization procedure reflect both the local and global properties of
background spacetime and the same holds for the vacuum state. In particular,
among the interesting directions in the investigations of the quantum vacuum
is the dependence of its properties on the spatial topology. Here we study
the polarization of the electromagnetic vacuum on background of $(D+1)$%
-dimensional de Sitter (dS) spacetime in the presence of a topological
defect that is a generalization of a straight cosmic string in
(3+1)-dimensional spacetime. The cosmic strings are formed in the early
universe as a result of symmetry breaking phase transitions (for reviews see 
\cite{Vile00,Hind95,Sake07,Cope10}). The interior energy density of those
defects is determined by the energy scale at which the phase transition
takes place and the observation of cosmic strings at recent epoch of the
universe expansion may provide an important window to high-energy physics.
The gravitational field around of those kinds of topological defects is a
source of a number of physical effects including the gravitational lensing,
the generation of gravitational waves and gamma ray bursts. The cosmic
strings also give rise to non-Gaussian distorsions in the fluctuations of
the cosmic microwave background \cite{Ring10}.

Our consideration of dS spacetime as a background geometry is motivated by
several reasons. First of all, its high degree of symmetry allows to obtain
closed analytic expressions for the cosmic string induced local
characteristics of the vacuum state. This will shed light on combined
effects of local geometry and topology in more complicated backgrounds.
Another reason is related to the profound role of dS spacetime in cosmology.
The most inflationary models employ the dS exponential expansion in the
early universe to provide natural solutions to a number of problems in the
Standard Big-Bang cosmology (see \cite{Lind94,Bass07,Mart14}). From the
other side, the observational data on the temperature anisotropies of cosmic
microwave background, high redshift supernovae and galaxy clusters \cite%
{Ries98}-\cite{Ade14} indicate that the recent expansion of the universe is
dominated by a source of the cosmological constant type. The relative
contribution of the latter to the total energy density will increase during
the expansion and the corresponding geometry asymptotes dS spacetime in the
future.

The quantum field-theoretical effects in fixed dS background were
investigated in a large number of papers. Different coordinate systems have
been considered, including global, planar, static and hyperbolic
coordinates. In particular, motivated by an important role of vacuum
fluctuations during the inflationary phase in the formation of large scale
structures in the post inflationary era, the dynamics of vacuum fluctuations
of scalar fields have attracted great deal of attention. The properties of
those fluctuations are encoded in the observed spectrum of temperature
anisotropies for cosmic microwave background radiation and this opens a
possibility to test the effects of gravity on quantum matter. A similar
amplification of the electromagnetic field quantum fluctuations by the dS
expansion can serve as a mechanism for the generation of extragalactic
magnetic fields. However, for this mechanism to work, it is required to
break the conformal invariance of the electromagnetic field in
(3+1)-dimensional spacetime. That can be done by an additional coupling of
the electromagnetic field with other degrees of freedom, for example, by
adding a factor in the Maxwell kinetic term that depends on other fields.
Another possibility for breaking the conformal symmetry is to consider
models with extra spatial dimensions (for the generation of large scale
magnetic fields in those types of models see, for example, \cite%
{Giov00,Giov04,Atmj14}). This is one of the reasons for our discussion of
background spacetime with general number of dimensions. Another reason is
that the background geometry we consider can also be generated by
fundamental strings in superstring theories. A mechanism for the formation
of those types of objects within the framework of brane inflation models has
been discussed in \cite{Ring10,Hind11,Cope11,Cher15}. In models with
fundamental strings as cosmic strings, depending upon the number of compact
dimensions, the spatial dimension of the effective theory may vary from 3 to
9.

The polarization of vacuum by a cosmic string has been widely considered in
the literature for both neutral and charged quantum fields. That was mainly
done for an idealized model where the influence of cosmic string on the
spacetime geometry outside the core is reduced to the generation of planar
angle deficit determined by the energy density inside the core. Exact
analytic results for the local characteristics of the vacuum are obtained
for highly symmetric background geometries, including Minkowski, dS and
anti-de Sitter (AdS) spacetimes. Cosmic string type topological defects in
dS and AdS spacetimes were considered in \cite{Ghez02}-\cite{Sant16}. The
influence of the cosmological expansion on the properties of scalar vacuum
around a straight cosmic string in flat Friedmann-Robertson-Walker type
models has been discussed in \cite{Davi88} for special values of the planar
angle deficit $2\pi -\phi _{0}$ corresponding to integer values of the
parameter $q=2\pi /\phi _{0}$. For those values the Green function in a
conical space is expressed as an image sum over the Green functions in the
absence of cosmic string. The vacuum polarization around a straight cosmic
string in dS spacetime for general values of the planar angle deficit in the
cases of massive scalar and fermionic fields has been investigated in \cite%
{Beze09,Beze10}. The correlators for the electromagnetic field and the
vacuum expectation values of the electric and magnetic fields squared were
discussed in \cite{Saha17,Saha18Part}. Additional topological effects on the
properties of the scalar and fermionic vacua induced by compactification of
a cosmic string along its axis in locally dS bulk were studied in \cite%
{Moha15,Brag20f,Brag20}. The polarization of the scalar and fermionic vacua
around a cosmic string in background of AdS spacetime has been investigated
in \cite{Beze12AdS,Beze13AdS}. The combined effects of a cosmic string,
compact dimensions and branes on the local characteristics of the vacuum in
locally AdS spacetime were discussed in \cite{Oliv19}-\cite{Bell22AdSc}. The
finite temperature effects on the expectation values of the charge and
current densities for a charged scalar field in the presence of a cosmic
string and compact dimension on locally AdS bulk were considered recently in 
\cite{Beze22AdSt}.

The organization of the present paper is as follows. The problem setup and
the mode functions for the electromagnetic field are presented in the
following section. Section \ref{sec:Diag} considers the vacuum expectation
values (VEVs) for the diagonal components of the energy-momentum tensor. The
contributions induced by the cosmic string are separated explicitly and
their asymptotics are investigated. Special cases are discussed and the
results of numerical analysis are presented. In addition to the diagonal
components, the vacuum energy-momentum tensor has a nonzero off-diagonal
component that describes energy flux along the radial direction. This is a
cosmic string induced effect and is studied in Section \ref{sec:Flux}. The
main results of the paper are summarized in Section \ref{sec:Conc}.

\section{Setup and electromagnetic mode functions}

\label{sec:Modes}

As a background geometry we consider $(D+1)$-dimensional spacetime having
the line element 
\begin{equation}
ds^{2}=dt^{2}-e^{2t/\alpha }[dr^{2}+r^{2}d\phi ^{2}+\left( d\mathbf{z}%
\right) ^{2}],  \label{ds2t}
\end{equation}%
where $\mathbf{z=}\left( z^{3},...,z^{D}\right) $ with $-\infty
<z^{i}<+\infty $, $-\infty <t<+\infty $, $0\leqslant r<\infty $, and $%
0\leqslant \phi \leqslant \phi _{0}$. In the special case $\phi _{0}=2\pi $
this line element describes $(D+1)$-dimensional dS spacetime sourced by the
cosmological constant $\Lambda =D(D-1)/(2\alpha ^{2})$ and foliated by flat
spaces with cylindrical coordinates $(r,\phi ,\mathbf{z})$. For $\phi
_{0}<2\pi $ and at the points $r>0$ the local geometrical characteristics
coincide with those for dS spacetime. In particular, for the Ricci scalar
one has $R=D(D+1)/\alpha ^{2}$. The geometry described by (\ref{ds2t}) is a
generalization of an idealized cosmic string, being a linear topological
defect in $D=3$, for spatial dimensions $D>3$. The defect core, given by $%
r=0 $, presents a $(D-2)$-dimensional spatial hypersurface. Introducing the
new time coordinate $\tau $ in accordance with $\tau =-\alpha e^{-t/\alpha }$%
, $-\infty <\tau <0$, the line element is written in the form conformally
flat for $r>0$: 
\begin{equation}
ds^{2}=\frac{\alpha ^{2}}{\tau ^{2}}[d\tau ^{2}-dr^{2}-r^{2}d\phi
^{2}-\left( d\mathbf{z}\right) ^{2}].  \label{ds2}
\end{equation}%
Similar to the case of the Minkowski bulk, for the model with $D=3$ the
deficit angle $2\pi -\phi _{0}$ in dS spacetime is induced by the vortex
solution of the Einstein-Abelian-Higgs equations in the presence of a
cosmological constant \cite{Ghez02}.

In quantum field theory the notion of the vacuum state, denoted as $%
|0\rangle $, has a global nature and its properties are sensitive to both
local and global characteristics of the background spacetime. In the
geometry under consideration, the nontrivial spatial topology induced by the
planar angle deficit gives rise to additional vacuum polarization for
quantum fields. Here we are interested in the topological effects on the VEV
of the energy-momentum tensor $T_{i}^{l}$ for the electromagnetic field with
the vector potential $A_{i}(x)$. Our approach for evaluation of the
expectation value $\langle 0|T_{i}^{l}|0\rangle \equiv \langle
T_{i}^{l}\rangle $ will be based on the direct summation over the complete
set of electromagnetic modes for the background geometry (\ref{ds2}). We
will denote by $\{A_{(\beta )i}(x),A_{(\beta )i}^{\ast }(x)\}$ the
corresponding set for the vector potential, where $\beta $ is the collective
notation for the quantum numbers specifying the solutions of Maxwell's
equations in $(D+1)$-dimensional spacetime. With $F_{(\beta )lm}=\partial
_{l}A_{(\beta )m}-\partial _{m}A_{(\beta )l}$ being the electromagnetic
field tensor for the modes, the mode-sum for the vacuum energy-momentum
tensor is expressed as%
\begin{equation}
\langle T_{i}^{l}\rangle =\frac{1}{4\pi }\underset{\beta }{\sum }\left[ 
\frac{1}{4}\delta _{i}^{l}F_{(\beta )}^{mn}F_{(\beta )mn}^{\ast }-F_{(\beta
)i}^{\cdot m}F_{(\beta )\cdot m}^{l\ast }\right] ,  \label{Tmu}
\end{equation}%
where $F_{(\beta )}^{il}=g^{im}g^{ln}F_{(\beta )mn}$. In what follows it is
convenient to work in the coordinates $(\tau ,r,\phi ,\mathbf{z})$ with the
metric tensor $g_{il}=\left( \alpha /\tau \right) ^{2}\mathrm{diag}%
(1,-1,-r^{2},-1,\ldots ,-1)$. We will fix the vector potential by the gauge
conditions $A_{0}=0$ and $\sum_{i=1}^{D}\partial _{i}(\sqrt{|g|}A^{i})=0$
with $\sqrt{|g|}=\left( \alpha /\eta \right) ^{D+1}r$ and $\eta =|\tau |$.

The electromagnetic cylindrical modes have been discussed in \cite{Saha16}
for $(D+1)$-dimensional dS spacetime and were generalized for the presence
of a cosmic string in \cite{Saha17}. One has $D-1$ polarization states and
we will distinguish them by the index $\sigma =1,\ldots ,D-1$. \ In addition
to this quantum number, the electromagnetic modes are specified by the wave
vector $\mathbf{k}=(k_{3},\ldots ,k_{D})$ in the subspace covered by the
coordinates $\mathbf{z}$, by the radial quantum number $\gamma $, $0\leq
\gamma <\infty $, and by the azimuthal quantum number $m$, $m=0,\pm 1,\pm
2,\ldots $. Consequently, the set $\beta $ is specified as $(\beta )=(\gamma
,m,\mathbf{k},\sigma )$. Introducing the notation $q=2\pi /\phi _{0}$, the
vector potentials for separate polarizations are expressed as \cite{Saha17}%
\begin{eqnarray}
A_{(\beta )l}(x) &=&\eta ^{\frac{D}{2}-1}Z_{\frac{D}{2}-1}(\omega \eta
)\left\{ 
\begin{array}{c}
\left( 0,iqm/r,-r\partial _{r},0,\ldots ,0\right) \\ 
\left( 0,\omega \epsilon _{\sigma n}+i\omega ^{-1}\mathbf{k}\cdot {%
\boldsymbol{\epsilon }}_{\sigma }\partial _{n}\right)%
\end{array}%
\right.  \notag \\
&&\times J_{q|m|}(\gamma r)e^{iqm\phi +i\mathbf{k}\cdot \mathbf{z}},\;%
\begin{array}{l}
\sigma =1 \\ 
\sigma =2,\ldots ,D-1%
\end{array}%
,  \label{Asig}
\end{eqnarray}%
where $\omega =\sqrt{\gamma ^{2}+|\mathbf{k}|^{2}}$, $n=1,\ldots ,D$, $%
J_{\nu }(y)$ is the Bessel function, $\mathbf{k}\cdot \mathbf{z}%
=\sum_{i=3}^{D}k_{i}z^{i}$, and $\mathbf{k}\cdot {\boldsymbol{\epsilon }}%
_{\sigma }=\sum_{i=3}^{D}k_{i}\epsilon _{\sigma i}$, with $\epsilon _{\sigma
1}=\epsilon _{\sigma 2}=0$. In (\ref{Asig}), the function $Z_{\nu }(y)$ is
the linear combination of the Hankel functions $H_{\nu }^{(1)}(y)$ and $%
H_{\nu }^{(2)}(y)$, $Z_{\nu }(y)=\sum_{i=1,2}c_{\beta }^{(i)}H_{\nu
}^{(i)}(y)$. The polarization vectors ${\boldsymbol{\epsilon }}_{\sigma }$
obey the orthonormalization condition%
\begin{equation}
\sum_{l,n=3}^{D}\left( \delta _{nl}-\frac{k_{l}k_{n}}{\omega ^{2}}\right)
\epsilon _{\sigma l}\epsilon _{\sigma ^{\prime }n}=\frac{\gamma ^{2}}{\omega
^{2}}\delta _{\sigma \sigma ^{\prime }}.  \label{Polrel1}
\end{equation}

The orthonormalization condition for the mode functions is written as 
\begin{equation}
\int d^{D}x\sqrt{|g|}[A_{(\beta ^{\prime })l}^{\ast }\nabla ^{0}A_{(\beta
)}^{l}-(\nabla ^{0}A_{(\beta ^{\prime })l}^{\ast })A_{(\beta )}^{l}]=4i\pi
\delta (\gamma -\gamma ^{\prime })\delta (\mathbf{k}-\mathbf{k}^{\prime
})\delta _{mm^{\prime }}\delta _{\sigma \sigma ^{\prime }},  \label{NC}
\end{equation}%
where $\nabla ^{i}=g^{ik}\nabla _{k}$ and $\nabla _{k}$ is the covariant
derivative operator. This condition determines the one of the coefficients $%
c_{\beta }^{(i)}$ in the linear combination of the Hankel functions in the
definition of the function $Z_{\nu }(y)$. Similar to that for pure dS bulk,
the second coefficient is fixed by the choice of the vacuum state. Here we
will assume that the electromagnetic field is prepared in the state that is
the analog of the Bunch-Davies vacuum for a scalar field in dS spacetime.
For that state $c_{\beta }^{(2)}=0$ and in the expressions (\ref{Asig}) for
the mode functions%
\begin{equation}
Z_{\frac{D}{2}-1}(\omega \eta )=c_{\beta }^{(1)}H_{\frac{D}{2}%
-1}^{(1)}(\omega \eta ).  \label{ZBD}
\end{equation}%
With this choice, the normalization coefficient is found from (\ref{NC}):%
\begin{equation}
|c_{\beta }^{(1)}|^{2}=\frac{q}{4(2\pi \alpha )^{D-3}\gamma }.  \label{cbet}
\end{equation}%
Note that this coefficient is the same for all the polarizations.

For $D=3$ one has $\eta ^{D/2-1}H_{D/2-1}^{(1)}(\omega \eta )=-ie^{-i\omega
\tau }\sqrt{2/\pi \omega }$ and the mode functions (\ref{Asig}) coincide
with those for a cosmic string in the Minkowski bulk. Of course, this is a
consequence of the conformal invariance of the electromagnetic field in
4-dimensional spacetime. For $D>3$ and in the limit $\alpha \rightarrow
\infty $ for a fixed value of the synchronous time $t$, from (\ref{Asig})
the mode functions are obtained for a cosmic string in $(D+1)$-dimensional
Minkowski bulk.

\section{Diagonal components of the energy-momentum tensor}

\label{sec:Diag}

Having the complete set of the electromagnetic modes we can evaluate the VEV
of the energy-momentum tensor by making use of the formula (\ref{Tmu}). The
expression in the right-hand side of (\ref{Tmu}) is divergent and a
regularization is required. For example, the point splitting procedure can
be employed with the combination of the two-point functions from \cite%
{Saha17} (for vector field correlators in dS spacetime see \cite{Alle86}-%
\cite{Glav22} and references therein). Another possibility is to introduce a
cutoff function, for example, in the form $e^{-b\omega ^{2}}$ with the
cutoff parameter $b>0$. The important point to be mentioned is the
following. For $r>0$ the local geometry in the presence of the cosmic string
is the same as that for the pure dS spacetime and, hence, for those points
the divergences in the VEV of the energy-momentum tensor are the same as
well. From here it follows that if one extracts explicitly the part in the
regularized VEV corresponding to the pure dS geometry, then the
renormalization is required for that part only. The remaining topological
contribution is finite in the limit when the regularization is removed.
Below we will extract the pure dS part and the specific regularization
scheme is not essential in that procedure. In the evaluation of the diagonal
components we will regularize the expectation values by introducing the
cutoff function $e^{-b\omega ^{2}}$.

\subsection{General expressions}

Substituting the modes (\ref{Asig}) in (\ref{Tmu}) with the cutoff function $%
e^{-b\omega ^{2}}$ and by using the relation \cite{Saha17}%
\begin{equation}
\sum_{\sigma =2}^{D-1}\epsilon _{\sigma n}\epsilon _{\sigma l}=\frac{\gamma
^{2}\delta _{nl}+k_{n}k_{l}}{\omega ^{2}},  \label{Polrel2}
\end{equation}%
with $l,n=3,...,D$, for summation over the polarization states, the
regularized diagonal components of the vacuum energy-momentum tensor are
expressed as (no summation over $i$) 
\begin{eqnarray}
\langle T_{i}^{i}\rangle _{\mathrm{reg}} &=&\frac{2^{2-D}q\eta ^{D+2}}{\pi
^{D/2+1}\Gamma (D/2-1)\alpha ^{D+1}}\sideset{}{'}{\sum}_{m=0}^{\infty
}\int_{0}^{\infty }dk\,k^{D-3}\int_{0}^{\infty }d\gamma \gamma e^{-b\omega
^{2}}  \notag \\
&&\times \sum_{l=1,2}t_{l}^{(i)}[k,\gamma ,J_{qm}(\gamma r)]K_{\frac{D}{2}%
-l}(\omega \eta e^{-i\pi /2})K_{\frac{D}{2}-l}(\omega \eta e^{i\pi /2}),
\label{Tii}
\end{eqnarray}%
where $k=|\mathbf{k}|$, $K_{\nu }(x)$ is the Macdonald function. The prime
on the sign of summation means that the term $m=0$ is taken with coefficient
1/2. The Macdonald function is introduced instead of the Hankel function in
the expression for the electromagnetic modes. In (\ref{Tii}), the function
containing the radial dependence is defined as%
\begin{equation}
t_{l}^{(i)}\left[ k,\gamma ,f\left( y\right) \right]
=(a_{l}^{(i)}k^{2}+b_{l}^{(i)}\gamma ^{2})\left[ f^{\prime 2}(y)+\delta _{i}%
\frac{q^{2}m^{2}}{y^{2}}f^{2}(y)\right] +\left[ \left( D-3\right)
c_{l}^{(i)}k^{2}+d_{l}^{(i)}\gamma ^{2}\right] f^{2}(y),  \label{til}
\end{equation}%
where 
\begin{equation}
\delta _{i}=\left\{ 
\begin{array}{ll}
1, & i=0,3,\ldots ,D \\ 
-1, & i=1,2%
\end{array}%
\right. .  \label{del1}
\end{equation}%
For the coefficients in the definition of the function $t_{l}^{(i)}\left[
k,\gamma ,f\left( y\right) \right] $ one has 
\begin{eqnarray}
a_{l}^{(0)} &=&\left( 2,2\right) ,\;b_{l}^{(0)}=\left( D-2,1\right)
,\;c_{l}^{(0)}=\left( 1,1\right) ,\;d_{l}^{(0)}=\left( 1,D-2\right) ,  \notag
\\
a_{l}^{(1)} &=&\left( 0,0\right) ,\;b_{l}^{(1)}=\left( 2-D,-1\right) ,\
c_{l}^{(1)}=\left( 1,-1\right) ,\;d_{l}^{(1)}=\left( -1,2-D\right) ,  \notag
\\
a_{l}^{(2)} &=&\left( 0,0\right) ,\;b_{l}^{(2)}=\left( D-2,1\right)
,\;c_{l}^{(2)}=\left( 1,-1\right) ,\;d_{l}^{(2)}=\left( -1,2-D\right) ,
\label{a2c2}
\end{eqnarray}%
and%
\begin{eqnarray}
a_{l}^{(p)} &=&\left( 2\frac{D-4}{D-2},-2\right) ,\;b_{l}^{(p)}=\left(
D-4,-1\right) ,  \notag \\
c_{l}^{(p)} &=&\left( \frac{D-6}{D-2},\frac{4-D}{D-2}\right)
,\;d_{l}^{(p)}=\left( 1,4-D\right) ,  \label{apcp}
\end{eqnarray}%
for $p=3,\ldots ,D$. Here, the first and second terms correspond to $l=1$
and $l=2$, respectively.

For the further transformation of (\ref{Tii}) we use the integral
representation \cite{Wats66}%
\begin{equation}
K_{\nu }(\omega \eta e^{-i\pi /2})K_{\nu }(\omega \eta e^{i\pi /2})=\frac{1}{%
2}\int_{0}^{\infty }\frac{dx}{x}\exp \left( -\omega ^{2}\frac{x}{4}+\eta ^{2}%
\frac{2}{x}\right) K_{\nu }\left( 2\eta ^{2}/x\right) ,  \label{Lint}
\end{equation}%
for the product of the Macdonald functions. Substituting in (\ref{Tii}),
after integration over $k$ we get (no summation over $i$) 
\begin{eqnarray}
\langle T_{i}^{i}\rangle _{\mathrm{reg}} &=&\frac{2^{-D}q\eta ^{D+2}}{\pi
^{D/2+1}\alpha ^{D+1}}\sideset{}{'}{\sum}_{m=0}^{\infty }\int_{0}^{\infty }%
\frac{dx}{x}u^{D/2}e^{2\eta ^{2}/x}\int_{0}^{\infty }d\gamma \gamma
e^{-\gamma ^{2}/u}  \notag \\
&&\times \sum_{l=1,2}t_{l}^{(i)}(\gamma ^{2}/u,J_{qm}(\gamma r))K_{\frac{D}{2%
}-l}(2\eta ^{2}/x),  \label{Tii2}
\end{eqnarray}%
where $u=4/(x+4b)$ and 
\begin{eqnarray}
t_{l}^{(i)}(v,f\left( y\right) ) &=&\left[ \left( \frac{D}{2}-1\right)
a_{l}^{(i)}+b_{l}^{(i)}v\right] \left[ f^{\prime 2}(y)+\delta _{i}\frac{%
q^{2}m^{2}}{y^{2}}f^{2}(y)\right]  \notag \\
&&+\left[ \left( D-3\right) \left( \frac{D}{2}-1\right)
c_{l}^{(i)}+d_{l}^{(i)}v\right] f^{2}(y).  \label{til2}
\end{eqnarray}%
The integrals over $\gamma $ are evaluated on the base of the formula \cite%
{Prud86} 
\begin{equation}
\int_{0}^{\infty }d\gamma \ \gamma e^{-\gamma ^{2}/u}J_{qm}^{2}(\gamma r)=%
\frac{u}{2}e^{-r^{2}u/2}I_{qm}(r^{2}u/2),  \label{IntBes}
\end{equation}%
with $I_{\nu }(x)$ being the modified Bessel function, and by using the
formulas presented in \cite{Saha17}. This gives (no summation over $i$)%
\begin{eqnarray}
\langle T_{i}^{i}\rangle _{\mathrm{reg}} &=&\frac{(\eta /r)^{D+2}q}{%
2^{D/2}\pi ^{D/2+1}\alpha ^{D+1}}\sideset{}{'}{\sum}_{m=0}^{\infty
}\int_{0}^{\infty }\frac{dx}{x}w^{D/2+1}e^{2\eta ^{2}/x}  \notag \\
&&\times \sum_{l=1,2}K_{\frac{D}{2}-l}(2\eta ^{2}/x)\hat{W}_{l}^{(i)}\left(
e^{-w}I_{qm}(w)\right) ,  \label{Tii4}
\end{eqnarray}%
where $w=2r^{2}/(x+4b)$ and the operators $\hat{W}_{l}^{(i)}$ are defined by
the expression

\begin{equation}
\hat{W}_{l}^{(i)}=b_{l}^{(i)}w\partial _{w}^{2}+\left[
(D/2-1)a_{l}^{(i)}+b_{l}^{(i)}(1+w)+d_{l}^{(i)}w\right] \partial
_{w}+e_{l}^{(i)},  \label{Wi1}
\end{equation}%
for the components with $i=0,3,...,D,$ and\bigskip\ by%
\begin{equation}
\hat{W}_{l}^{(i)}=(d_{l}^{(i)}-b_{l}^{(i)})w\partial _{w}+e_{l}^{(i)},
\label{Wi2}
\end{equation}%
for $i=1,2$. Here the notations%
\begin{equation}
e_{l}^{(0)}=\frac{D-1}{2}\left( D,D\right) ,\;e_{l}^{(i)}=\frac{D-1}{2}%
\left( D-4,2-D\right) ,  \label{eli}
\end{equation}%
are introduced with $i=1,2,\ldots ,D$.

Next, in order to separate the topological contribution, we use the formula 
\cite{Beze10b}%
\begin{equation}
q\sideset{}{'}{\sum}_{m=0}^{\infty }e^{-w}I_{qm}\left( w\right) =%
\sideset{}{'}{\sum}_{j=0}^{[q/2]}e^{-2ws_{j}^{2}}-\frac{q}{\pi }\sin (q\pi
)\int_{0}^{\infty }dz\frac{e^{-2w\cosh ^{2}z}}{\cosh (2qz)-\cos (q\pi )},
\label{Sumj}
\end{equation}%
where $s_{j}=\sin (j\pi /q)$ and $[q/2]$ is the integer part of $q/2$. The
prime on the summation sign in the right-hand side means that the terms $j=0$
and $j=q/2$ (for even values of $q$) should be taken with additional
coefficient $1/2$. In the case $q=1$ the $j=0$ term remains only. This shows
that the part in the VEV corresponding to the $j=0$ term in (\ref{Sumj})
corresponds to the regularized VEV of the energy-momentum tensor in dS
spacetime in the absence of the cosmic string. It is presented as (no
summation over $i$)%
\begin{equation}
\langle T_{i}^{i}\rangle _{\mathrm{reg}}^{\mathrm{(dS)}}=\frac{\alpha ^{-D-1}%
}{\left( 2\pi \right) ^{D/2+1}}\int_{0}^{\infty }dy\frac{y^{D/2}e^{y}}{%
(1+2by/\eta ^{2})^{D/2+1}}\sum_{l=1,2}K_{\frac{D}{2}-l}(y)e_{l}^{(i)}.
\label{TiidS}
\end{equation}%
The renormalized VEV of the energy-momentum tensor in dS spacetime, $\langle
T_{i}^{l}\rangle _{\mathrm{ren}}^{\mathrm{(dS)}}$, has been widely
investigated in the literature and our main concern here will be the cosmic
string induced contribution. From the maximal symmetry of the dS spacetime
and of the Bunch-Davies vacuum state the structure $\langle T_{i}^{l}\rangle
_{\mathrm{ren}}^{\mathrm{(dS)}}=\mathrm{const}\cdot \delta _{i}^{l}$ is
expected for the geometry in the absence of the cosmic string.

For the topological part induced by the cosmic string one has (no summation
over $i$)%
\begin{equation}
\langle T_{i}^{i}\rangle _{\mathrm{t}}=\underset{b\rightarrow 0}{\lim }\left[
\langle T_{i}^{i}\rangle _{\mathrm{reg}}-\langle T_{i}^{i}\rangle _{\mathrm{%
reg}}^{\mathrm{(dS)}}\right] .  \label{Tiit}
\end{equation}%
For $r>0$ the limit in the right-hand side is finite. From (\ref{Tii4}), by
using (\ref{Sumj}), we find the following representation (no summation over $%
i$)%
\begin{equation}
\langle T_{i}^{i}\rangle _{\mathrm{t}}=\frac{2\alpha ^{-D-1}}{(2\pi )^{D/2+1}%
}\left[ \sideset{}{'}{\sum}_{j=1}^{[q/2]}t^{(i)}\left( r/\eta ,s_{j}\right) -%
\frac{q}{\pi }\sin (q\pi )\int_{0}^{\infty }dz\frac{t^{(i)}\left( r/\eta
,\cosh z\right) }{\cosh (2qz)-\cos (q\pi )}\right] .  \label{Tiit1}
\end{equation}%
In (\ref{Tiit1}) we have used the notation%
\begin{equation}
t^{(i)}(x,y)=\int_{0}^{\infty }duu^{\frac{D}{2}}e^{u-2ux^{2}y^{2}}%
\sum_{l=1,2}K_{\frac{D}{2}-l}(u)f_{l}^{(i)}(x,y,u),  \label{ti}
\end{equation}%
with the functions%
\begin{equation}
f_{l}^{(i)}(x,y,u)=\left[
4b_{l}^{(i)}ux^{2}y^{2}-2(b_{l}^{(i)}+d_{l}^{(i)})ux^{2}-((D-2)a_{l}^{(i)}+2b_{l}^{(i)})%
\right] y^{2}+e_{l}^{(i)},  \label{tilt}
\end{equation}%
for $i=0,3,...,D$, and%
\begin{equation}
f_{l}^{(i)}(x,y,u)=2(b_{l}^{(i)}-d_{l}^{(i)})ux^{2}y^{2}+e_{l}^{(i)},
\label{tilt2}
\end{equation}%
for $i=1,2$. The topological part of the VEV depends on $\eta $ and $r$ in
the form of the ratio $r/\eta $. This property is a consequence of the
maximal symmetry of dS spacetime. By considering that $\alpha r/\eta $ is
the proper distance from the string, we see that $r/\eta $ is the proper
distance, measured in units of the dS curvature scale $\alpha $.

Note that the function (\ref{ti}) can also be written in the form%
\begin{eqnarray}
t^{(i)}(x,y) &=&2\int_{0}^{\infty }du\,u^{\frac{D}{2}}e^{u-2ux^{2}y^{2}}%
\sum_{l=1,2}K_{\frac{D}{2}-l}(u)  \notag \\
&&\times \left[
2a_{li}^{(4)}ux^{2}y^{4}+a_{li}^{(3)}ux^{2}y^{2}+a_{li}^{(2)}y^{2}+\frac{D-1%
}{4}a_{li}^{(1)}\right] ,  \label{ti2}
\end{eqnarray}%
with $i=0,1,2,3$. The coefficients are given by the expressions%
\begin{eqnarray}
a_{li}^{(1)} &=&\left( 
\begin{array}{cccc}
D & D-4 & D-4 & D-4 \\ 
D & 2-D & 2-D & 2-D%
\end{array}%
\right) ,  \notag \\
a_{li}^{(2)} &=&\left( 
\begin{array}{cccc}
2\left( 2-D\right) & 0 & 0 & 2(4-D) \\ 
1-D & 0 & 0 & D-1%
\end{array}%
\right) ,  \label{a2}
\end{eqnarray}%
and 
\begin{eqnarray}
a_{li}^{(3)} &=&\left( 
\begin{array}{cccc}
1-D & 3-D & D-1 & 3-D \\ 
1-D & D-3 & D-1 & D-3%
\end{array}%
\right) ,  \notag \\
a_{li}^{(4)} &=&\left( 
\begin{array}{cccc}
D-2 & 0 & 0 & D-4 \\ 
1 & 0 & 0 & -1%
\end{array}%
\right) .  \label{a4}
\end{eqnarray}%
For odd values of $D$ the integral in (\ref{ti2}) is expressed in terms of
elementary functions. By analogy with an ideal fluid, the diagonal
components $\langle T_{i}^{i}\rangle _{\mathrm{t}}$ of the energy-momentum
tensor can be interpreted as topological contributions to the vacuum
pressures along respective directions, $P_{\mathrm{t}i}=-\langle
T_{i}^{i}\rangle _{\mathrm{t}}$ with $i=1,2,\ldots ,D$. These pressures are
anisotropic.

It can be checked that the diagonal components obey the trace relation%
\begin{equation}
\langle T_{i}^{i}\rangle _{\mathrm{t}}=(3-D)\langle L\rangle _{\mathrm{t}},
\label{Trace}
\end{equation}%
where the topological part in the VEV of the Lagrangian density $%
L=-F_{il}F^{il}/(16\pi )$ is given by%
\begin{equation}
\langle L\rangle _{\mathrm{t}}=\frac{2\alpha ^{-D-1}}{(2\pi )^{D/2+1}}\left[
\sum_{l=1}^{[q/2]}g_{L}(r/\eta ,s_{l})-\frac{q}{\pi }\sin (q\pi
)\int_{0}^{\infty }dy\frac{g_{L}(r/\eta ,\cosh y)}{\cosh (2qy)-\cos (q\pi )}%
\right] ,  \label{Lt}
\end{equation}%
with the function%
\begin{eqnarray}
g_{L}(x,y) &=&\int_{0}^{\infty }du\,u^{\frac{D}{2}}e^{u-2x^{2}y^{2}u}\left\{
K_{\frac{D}{2}-2}(u)\left[ 2ux^{2}y^{2}\left( 2y^{2}-D+1\right) +\left(
D-1\right) \left( D/2-2y^{2}\right) \right] \right.  \notag \\
&&-\left. K_{\frac{D}{2}-1}(u)\left[ 2ux^{2}y^{2}\left(
2(D-2)y^{2}-D+1\right) +(D-1)D/2-4(D-2)y^{2}\right] \right\} .  \label{gL}
\end{eqnarray}%
The electric and magnetic contributions to the VEV of the Lagrangian density
have been discussed in \cite{Saha17}. By using the corresponding expressions
for the VEVs of the squared electric and magnetic fields, we can check that
the formula (\ref{Lt}) is obtained. In the special case $D=3$ the
electromagnetic field is conformally invariant and the topological part is
traceless. The trace relation $T_{i}^{i}=(3-D)L$ in classical
electrodynamics directly follows from the definition of the energy-momentum
tensor. In quantum field theory anomalies may arise in the corresponding
relation between the VEVs. The well-known example is the trace anomaly for
the energy-momentum tensor of a conformally coupled massless scalar field.
In the problem at hand we have local dS geometry for $r>0$ and the trace
anomaly is contained in the pure dS part of the VEV. The trace relation for
the topological parts in the VEVs is the same as that in classical
electrodynamics.

\subsection{Special cases and asymptotics}

In the special cases $D=3$ and $D=5$ from (\ref{ti}) one gets%
\begin{eqnarray}
t^{(i)}(x,y) &=&-\frac{\sqrt{2\pi }}{4(xy)^{4}}A^{(i)},\ D=3,  \notag \\
t^{(i)}(x,y) &=&\frac{\sqrt{2\pi }}{4(xy)^{6}}%
(B^{(i)}y^{2}+C^{(i)}x^{2}y^{2}+D^{(i)}),\ D=5,  \label{ti5}
\end{eqnarray}%
where 
\begin{eqnarray}
A^{(0)} &=&A^{(1)}=A^{(3)}=1,\ A^{(2)}=-3,  \notag \\
B^{(0)} &=&B^{(i)}=2,\ B^{(1)}=B^{(2)}=0,  \notag \\
C^{(0)} &=&-C^{(1)}=1,\ C^{(2)}=5,  \notag \\
D^{(0)} &=&D^{(1)}=-2,\ D^{(2)}=10,  \label{D01}
\end{eqnarray}%
and $C^{(i)}=-1$,$\;D^{(i)}=-2$, for $i=3,4,5$. The corresponding
topological parts in the energy-momentum tensor are expressed in terms of
the functions%
\begin{equation}
c_{n}(q)=\sum_{j=1}^{[q/2]}s_{j}^{-n}-\frac{q}{\pi }\sin (q\pi
)\int_{0}^{\infty }dz\frac{\cosh ^{-n}z}{\cosh (2qz)-\cos (q\pi )}.
\label{cn}
\end{equation}%
For even values of $n$, the functions $c_{n}(q)$ can be found by using the
recurrence scheme described in \cite{Beze06}. In particular, one has $%
c_{2}(q)=(q^{2}-1)/6$ and%
\begin{eqnarray}
c_{4}(q) &=&\frac{q^{2}-1}{90}(q^{2}+11),  \notag \\
c_{6}(q) &=&\frac{q^{2}-1}{1890}(2q^{4}+23q^{2}+191).  \label{c6q}
\end{eqnarray}%
By using these results we find (no summation over $i$)%
\begin{equation}
\langle T_{i}^{i}\rangle _{\mathrm{t}}=-\frac{A^{(i)}c_{4}(q)}{8\pi
^{2}(\alpha r/\eta )^{4}},  \label{TiiD3}
\end{equation}%
for $D=3$ and%
\begin{equation}
\langle T_{i}^{i}\rangle _{\mathrm{t}}=\frac{\left[ B^{(i)}+(r/\eta
)^{2}C^{(i)}\right] c_{4}(q)+D^{(i)}c_{6}(q)}{16\pi ^{3}(\alpha r/\eta )^{6}}%
,  \label{TiiD5}
\end{equation}%
for $D=5$. In the case $D=3$ the off-diagonal components of the
energy-momentum tensor vanish (see below) and from (\ref{TiiD3}) one gets
(no summation over $i$) $\langle T_{i}^{i}\rangle _{\mathrm{t}}=(\eta
/\alpha )^{4}\langle T_{i}^{i}\rangle _{\mathrm{t}}^{\mathrm{(M)}}$, where $%
\langle T_{i}^{i}\rangle _{\mathrm{t}}^{\mathrm{(M)}}$ is the corresponding
VEV in Minkowski bulk \cite{Frol87,Dowk87}. This is a direct consequence of
the conformal invariance of the electromagnetic field in $D=3$ spatial
dimensions and of the conformal flatness of the background geometry. For $%
D=3 $ the energy density is negative everywhere, whereas for $D=5$ it can be
either negative or positive, depending on the values of the parameters (see
the graphs below).

In a similar way, for $D=7$ we get (no summation over $i$)%
\begin{equation}
\langle T_{i}^{i}\rangle _{\mathrm{t}}=\frac{3\pi ^{-4}}{2^{6}\left( \alpha
r/\eta \right) ^{8}}\sum_{l=1}^{3}E_{i+1,l}(r/\eta )c_{2+2l}(q),
\label{TiiD7}
\end{equation}%
with the $4\times 3$ matrix%
\begin{equation}
E_{k,l}(x)=\left( 
\begin{array}{ccc}
x^{2}(8+9x^{2}) & 4(2+x^{2}) & -3 \\ 
x^{4} & -4x^{2} & -3 \\ 
21x^{4} & 28x^{2} & 21 \\ 
x^{2}(8+x^{2}) & 4(2-x^{2}) & -3%
\end{array}%
\right) ,  \label{Ekl}
\end{equation}%
where $k=1,2,3,4$ and $l=1,2,3$. The functions $c_{4}(q)$ and $c_{6}(q)$ are
given by (\ref{c6q}) and the function $c_{8}(q)$ is obtained by using the
recurrence scheme from \cite{Beze06}:%
\begin{equation}
c_{8}(q)=\frac{q^{2}-1}{28350}\left( 3q^{6}+43q^{4}+337q^{2}+2497\right) .
\label{c8q}
\end{equation}%
As it will be seen below, the functions $c_{n}(q)$ also appear in the
asymptotic expressions for the components of the energy-momentum tensor. The
function $c_{n}(q)$, $n\geq 1$, is positive for $q>1$ and monotonically
increasing in that region. Additionally, one has $c_{n}(1)=0$ and $%
c_{n}(2)=1/2$. We also have the properties $c_{n+1}(q)<c_{n}(q)$ for $1<q<2$
and $c_{n}(q)<c_{n+1}(q)$ for $q>2$. The asymptotic behavior of the function 
$c_{n}(q)$ for large values of $q$ can be found from (\ref{cn}). For $q\gg 1$
the dominant contribution in (\ref{cn}) comes from the terms in the sum with 
$j\ll q$ and we get $c_{n}(q)\approx \left( q/\pi \right) ^{n}\zeta (n)$,
where $\zeta (n)$ is the Riemann zeta function. Note that $\zeta (4)=\pi
^{4}/90$, $\zeta (6)=\pi ^{6}/945$, $\zeta (8)=\pi ^{8}/9450$, and this
estimate is in agreement with the exact formulas (\ref{c6q}) and (\ref{c8q}).

The Minkowskian limit corresponds to $\alpha \rightarrow \infty $ for a
fixed value of the time coordinate $t$. In this case one has $\eta \approx
\alpha -t$ and $\eta $ is large. Hence, we need the asymptotic of the
function $t^{(i)}(x,y)$ for small values of $x$. In this limit the dominant
contribution to the integral in (\ref{ti}) comes from large values of $u$
and using the asymptotic expression $K_{\nu }(u)\approx \sqrt{\pi /(2u)}%
e^{-u}$ for the Macdonald function for large argument, to the leading order
we get the VEV $\langle T_{i}^{k}\rangle _{\mathrm{t}}^{\mathrm{(M)}}$ for a
cosmic string on the Minkowski bulk. The VEV is diagonal and the expression
for the components with $i=0,3,\ldots ,D$ has the form (no summation over $i$%
) \cite{Saha18Part}%
\begin{equation}
\langle T_{i}^{i}\rangle _{\mathrm{t}}^{\mathrm{(M)}}(r)=\frac{\Gamma
(\left( D+1\right) /2)}{\left( 4\pi \right) ^{(D+1)/2}r^{D+1}}\left[ \left(
D-3\right) ^{2}c_{D-1}(q)-\left( D-1\right) c_{D+1}(q)\right] .  \label{TllM}
\end{equation}%
The corresponding energy density can be either positive or negative,
depending on the parameters $q$ and $D$. The radial and azimuthal components
are given by the expressions 
\begin{equation}
\langle T_{1}^{1}\rangle _{\mathrm{t}}^{\mathrm{(M)}}(r)=-\frac{1}{D}\langle
T_{2}^{2}\rangle _{\mathrm{t}}^{\mathrm{(M)}}(r)=-\frac{\Gamma (\left(
D+1\right) /2)}{\left( 4\pi \right) ^{(D+1)/2}r^{D+1}}\left( D-1\right)
c_{D+1}(q).  \label{T11M}
\end{equation}%
In the special case $D=3$ from (\ref{TllM}) and (\ref{T11M}) we get the
result obtained in \cite{Frol87,Dowk87}. For the Minkowski bulk, the
stresses $\langle T_{l}^{l}\rangle _{\mathrm{t}}^{\mathrm{(M)}}(r)$ along
the directions $l=3,\ldots ,D$ are equal to the energy density $\langle
T_{0}^{0}\rangle _{\mathrm{t}}^{\mathrm{(M)}}(r)$. The corresponding
equation of state, $P_{\mathrm{t}l}^{\mathrm{(M)}}=-\langle T_{l}^{l}\rangle
_{\mathrm{t}}^{\mathrm{(M)}}=-\langle T_{0}^{0}\rangle _{\mathrm{t}}^{%
\mathrm{(M)}}$, is of the cosmological constant type, though the energy
density and pressures depend on the radial coordinate. For dS background
geometry with spatial dimensions $D>3$ the stresses along the directions
parallel to the core differ from the energy density.The energy density
corresponding to (\ref{TllM}) is negative for $D=3,4$. In spatial dimensions 
$D\geq 5$ the energy density becomes zero for special values $q_{0}$ of the
parameter $q$. In those dimensions one has $\langle T_{0}^{0}\rangle _{%
\mathrm{t}}^{\mathrm{(M)}}>0$ for $1<q<q_{0}$ and $\langle T_{0}^{0}\rangle
_{\mathrm{t}}^{\mathrm{(M)}}>0$ for $q>q_{0}$. The critical value $q_{0}$
increases with increasing $D$. For example, we have $q_{0}=2$ for $D=5$ and $%
q_{0}\approx 3.83,4.86$ for $D=6,7$, respectively. For large values of $q$,
by using the corresponding asymptotic for the function $c_{n}(q)$, we get
(no summation over $i$)%
\begin{equation}
\langle T_{i}^{i}\rangle _{\mathrm{t}}^{\mathrm{(M)}}(r)\approx \frac{\left(
1-D\right) \zeta (D+1)}{\pi ^{3(D+1)/2}(2r)^{D+1}}\Gamma \left( \frac{D+1}{2}%
\right) q^{D+1},  \label{TiiMlq}
\end{equation}%
for $i=0,1,3,\ldots ,D$.

Now let us consider the asymptotic behavior of the diagonal components for
the energy-momentum tensor at large and small distances from the cosmic
string. At small proper distances compared to the curvature radius of the
background spacetime one has $r/\eta \ll 1$ and we need the asymptotic of
the function $t^{(i)}(x,y)$ for $x\ll 1$. For those $x$ the dominant
contribution to the integral in (\ref{ti}) comes from large values of $u$.
By using the corresponding asymptotic for the Macdonald function we get%
\begin{equation}
t^{(i)}(x,y)\approx \frac{\sqrt{\pi }\Gamma \left( (D+1)/2\right) }{%
2^{D/2+1}\left( xy\right) ^{D+1}}t^{(i)}(y),  \label{tias}
\end{equation}%
where $t^{(i)}(y)=\left( D-3\right) ^{2}y^{2}-D+1$ for $i=0,3,\ldots ,D$,
and $t^{(1)}(y)=-t^{(2)}(y)/D=1-D$. Substituting those asymptotics in (\ref%
{Tiit1}) we see that, to the leading order (no summation over $i$), 
\begin{equation}
\langle T_{i}^{i}\rangle _{\mathrm{t}}\approx \langle T_{i}^{i}\rangle _{%
\mathrm{t}}^{\mathrm{(M)}}(\alpha r/\eta ),\;r/\eta \ll 1.  \label{Tiinear}
\end{equation}%
Hence, the leading terms in the asymptotic expansions of the diagonal
components near the string coincide with the corresponding expressions for
the Minkowski bulk with the distance from the string replaced by the proper
distance in the dS bulk. In particular, the energy density near the cosmic
string is negative in spatial dimensions $D=3,4$. As it has been discussed
above, for $D\geq 5$ for critical values $q=q_{0}$ the energy density $%
\langle T_{0}^{0}\rangle _{\mathrm{t}}^{\mathrm{(M)}}$ becomes zero. Near
those points and for dS bulk, in the expansion over $\alpha r/\eta $ the
next to the leading terms should be kept. For large values of $q$ the energy
density near the cosmic string is negative with the asymptotic obtained from
(\ref{TiiMlq}) making the replacement $r\rightarrow \alpha r/\eta $. Near
the cosmic string the radial stress is negative and the azimuthal stress is
positive, $\langle T_{1}^{1}\rangle _{\mathrm{t}}<0$ and $\langle
T_{2}^{2}\rangle _{\mathrm{t}}>0$.

At large distances from the cosmic string one has $r/\eta \gg 1$ and in (\ref%
{Tiit1}) the asymptotic for the function $t^{(i)}(x,y)$ is required for
large values of the first argument. For $D=3$ we have a conformal relation
with the VEV in the Minkowski bulk and $\langle T_{i}^{i}\rangle _{\mathrm{t}%
}\propto 1/(r/\eta )^{4}$ for all values of the ratio $r/\eta $. In order to
estimate the integral for $D\geq 4$, we note that in the limit under
consideration the dominant contribution to the integral in (\ref{ti}) comes
from the region near the lower limit and we use the asymptotic expression
for the Macdonald function for small argument. In the leading order, this
gives%
\begin{equation}
t^{(i)}(x,y)\approx 2^{D/2-1}\frac{\Gamma (D/2-1)}{16(xy)^{4}}t_{0}^{(i)},
\label{tias2}
\end{equation}%
with the notations%
\begin{eqnarray}
t_{0}^{(0)} &=&(D-1)(D-4),\;t_{0}^{(2)}=D(D-1),  \notag \\
t_{0}^{(i)} &=&(D-3)(D-6)-2,\;i=1,3,...,D.  \label{ti0}
\end{eqnarray}%
The exception is the case $D=4$ for the function $t^{(0)}(x,y)$:%
\begin{equation}
t^{(0)}(x,y)\approx -\frac{3}{2}\frac{\ln (xy)}{(xy)^{6}},\;D=4.
\label{TiasD4}
\end{equation}%
With the asymptotic (\ref{tias2}), from (\ref{Tiit1}) we find the leading
behavior at large distances (no summation over $i$) 
\begin{equation}
\langle T_{i}^{i}\rangle _{\mathrm{t}}\approx \frac{\Gamma
(D/2-1)t_{0}^{(i)}c_{4}(q)}{32\pi ^{D/2+1}\alpha ^{D+1}(r/\eta )^{4}}%
,\;D\geq 4,  \label{Tiitlarge}
\end{equation}%
except $i=0$ for $D=4$. For the energy density in the special case $D=4$ one
gets%
\begin{equation}
\langle T_{0}^{0}\rangle _{\mathrm{t}}\approx -\frac{3c_{6}(q)\ln (r/\eta )}{%
8\pi ^{3}\alpha ^{4}(r/\eta )^{6}},\;D=4,  \label{T00tlarge}
\end{equation}%
and the decay is stronger. At large distance from the cosmic string the
topological contribution to the energy density is negative for $D=3,4$ and
positive for $D>4$. The stresses $\langle T_{i}^{i}\rangle _{\mathrm{t}}$,
with $i=1,3,...,D$, are negative for $3\leq D\leq 6$ and positive for $D>6$.
The stress $\langle T_{2}^{2}\rangle _{\mathrm{t}}$ is positive for $D\geq 3$%
. It is of interest to note that the topological contributions in the
diagonal components decay at large distances as the inverse fourth power of
the proper distance from the cosmic string in all spatial dimensions $D\geq
3 $. The exception is the energy density in 4-dimensional space with the
leading term (\ref{T00tlarge}). This behavior is in contrast to the geometry
of a cosmic string in the Minkowski bulk where the VEV decays like $%
1/r^{D+1} $.

Finally, we turn to the asymptotic for large values of the parameter $q$ and
fixed $r/\alpha $, $q\gg 1,r/\alpha $. The dominant contribution to the
topological parts (\ref{Tiit1}) come from the terms in the sum over $j$ with 
$j\pi /q\ll 1$. For those $j$ one has $s_{j}\approx $ $j\pi /q\ll 1$ and we
need the asymptotic of the function $t^{(i)}(x,y)$ for $y\ll 1$ and $xy\ll 1$%
. In the region under consideration the integral in (\ref{ti2}) is dominated
by the contribution from large values of $u$. By using the corresponding
asymptotic for the function $K_{D/2-l}(u)$, to the leading order we get%
\begin{equation}
t^{(i)}(x,y)\approx \frac{\sqrt{\pi }\left( 1-D\right) }{2^{D/2+1}\left(
xy\right) ^{D+1}}\Gamma \left( \frac{D+1}{2}\right) \left[ A^{(i)}-\delta
_{2}^{i}(D-3)\right] ,  \label{tilq}
\end{equation}%
for $i=0,1,2,3$. Substituting this, with $y=s_{j}\approx $ $j\pi /q$, in (%
\ref{Tiit1}), the sum over $j$ is approximated by the Riemann zeta function $%
\zeta (D+1)$ and for the leading term one finds (no summation over $i$)%
\begin{equation}
\langle T_{i}^{i}\rangle _{\mathrm{t}}\approx \frac{\left( 1-D\right) \zeta
(D+1)}{\pi ^{3(D+1)/2}\left( 2\alpha r/\eta \right) ^{D+1}}\Gamma \left( 
\frac{D+1}{2}\right) \left[ A^{(i)}-\delta _{2}^{i}(D-3)\right] q^{D+1},
\label{Tiilq}
\end{equation}%
for $q\gg 1$. We can show that in the special cases $D=3,5,7$ this general
result is in agreement with (\ref{TiiD3}), (\ref{TiiD5}), and (\ref{TiiD7}).
For $D=3$ the estimate (\ref{Tiilq}) directly follows from (\ref{TiiD3})
with combination of the approximation $c_{n}(q)\approx \left( q/\pi \right)
^{n}\zeta (n)$ for large $q$. For $D=5$ and $D=7$ we note that the dominant
contributions to the corresponding expressions come from the terms
containing $c_{6}(q)$ and $c_{8}(q)$, respectively, and the same
approximation for those functions gives the result (\ref{Tiilq}). The
leading term (\ref{Tiilq}) is traceless and the stresses $\langle
T_{i}^{i}\rangle _{\mathrm{t}}$ with $i=1,3,\ldots ,D$, coincide with the
energy density $\langle T_{0}^{0}\rangle _{\mathrm{t}}$. The latter is
negative for all values of $D$. In the Minkowskian limit the estimate agrees
with (\ref{TiiMlq}).

\begin{figure}[tbph]
\begin{center}
\begin{tabular}{cc}
\epsfig{figure=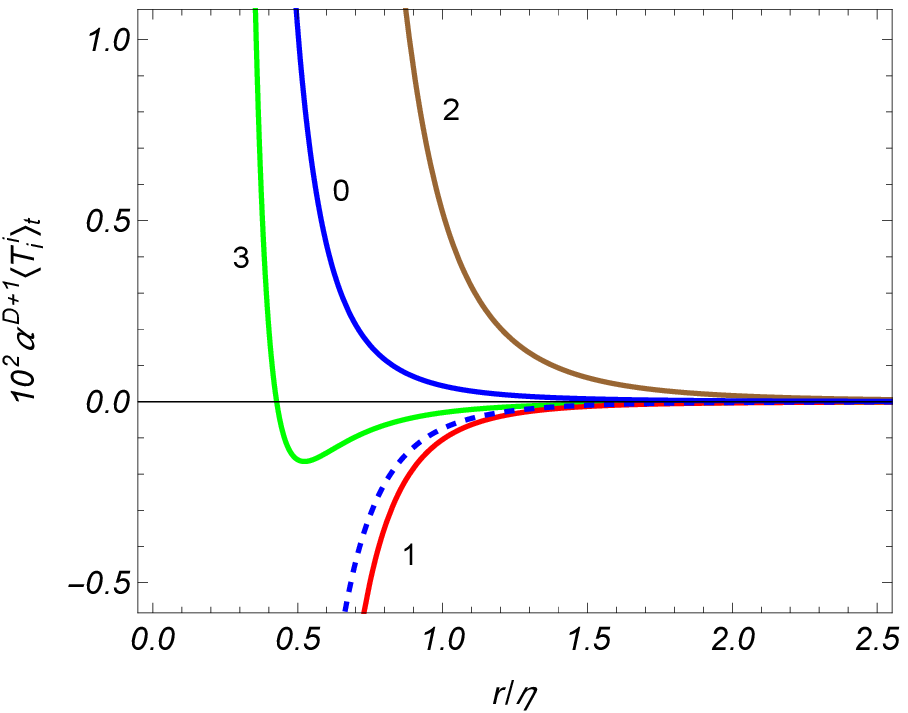,width=7.cm,height=6.cm} & \quad %
\epsfig{figure=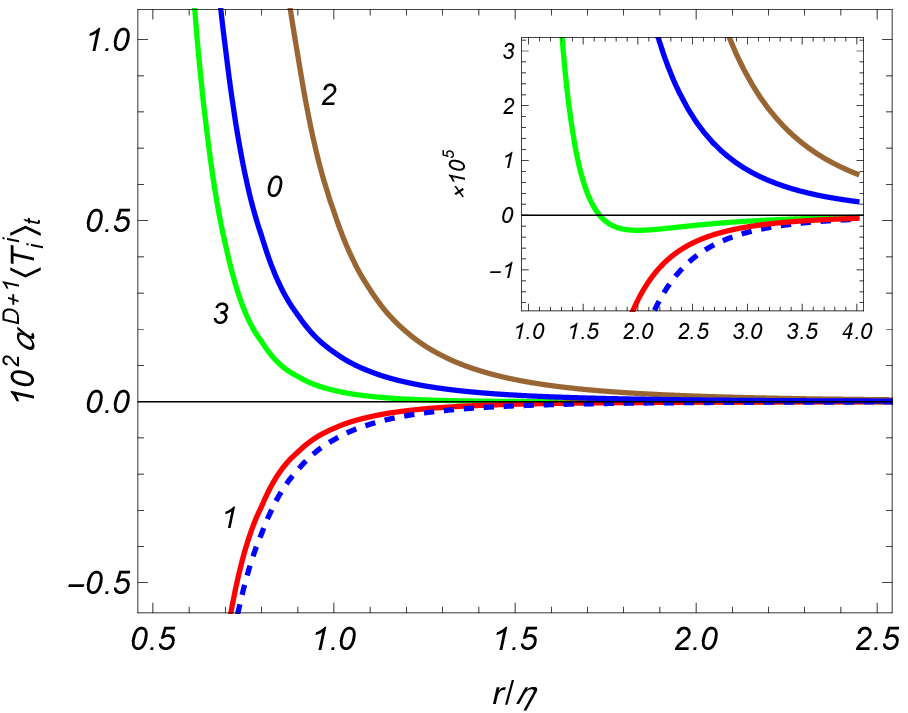,width=7.cm,height=6.cm}%
\end{tabular}%
\end{center}
\caption{The radial dependence of the topological contributions in the
vacuum energy-momentum tensor for spatial dimensions $D=5$ (left panel) and $%
D=6$ (right panel). The numbers near the curves correspond to the value of
the index $i$ of the diagonal components $\langle T_{i}^{i}\rangle _{\mathrm{%
t}}$ and the graphs are plotted for $q=1.5$. The dashed curves present the
energy flux $\langle T_{0}^{1}\rangle _{\mathrm{t}}$.}
\label{fig1}
\end{figure}

In Figure \ref{fig1} we have displayed the dependence of the diagonal
components of the topological contributions in the VEV of the
energy-momentum tensor, $\langle T_{i}^{i}\rangle _{\mathrm{t}}$ (in units
of $1/\alpha ^{D+1}$), on the ratio $r/\eta $ (proper distance from the
cosmic string in units of the dS curvature radius $\alpha $). The numbers
near the curves correspond to the value of the index $i$ and the left and
right panels are plotted for $D=5$ and $D=6$, respectively. For the planar
angle deficit we have taken the value corresponding to $q=1.5$. The dashed
curves on both panels present the energy flux (see below). For both cases $%
D=5,6$ the energy density is positive. For $D=3,4$ the corresponding energy
density is negative. In general, as it has been demonstrated by the
asymptotic analysis, depending on the values of $D$ and $q$, the energy
density can be either positive or negative. For the values of the parameters
corresponding to Figure \ref{fig1} the radial and azimuthal stresses are
monotonic functions of $r/\eta $, whereas the stresses $\langle
T_{i}^{i}\rangle _{\mathrm{t}}$, $i=1,3,\ldots ,D$, are positive near the
cosmic string and negative al large distances. 

\begin{figure}[tbph]
\begin{center}
\begin{tabular}{cc}
\epsfig{figure=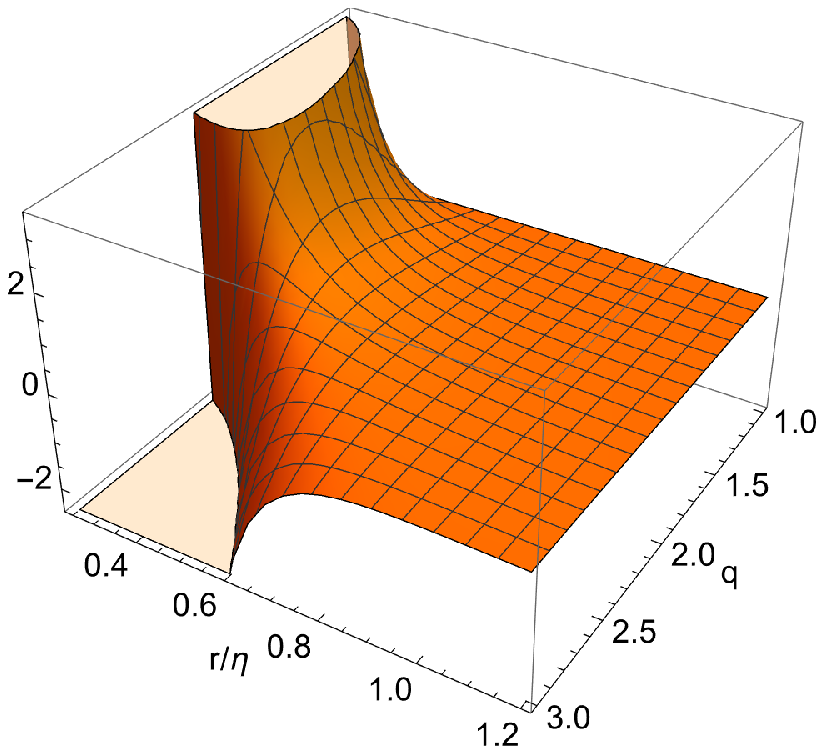,width=7.cm,height=6.cm} & \quad %
\epsfig{figure=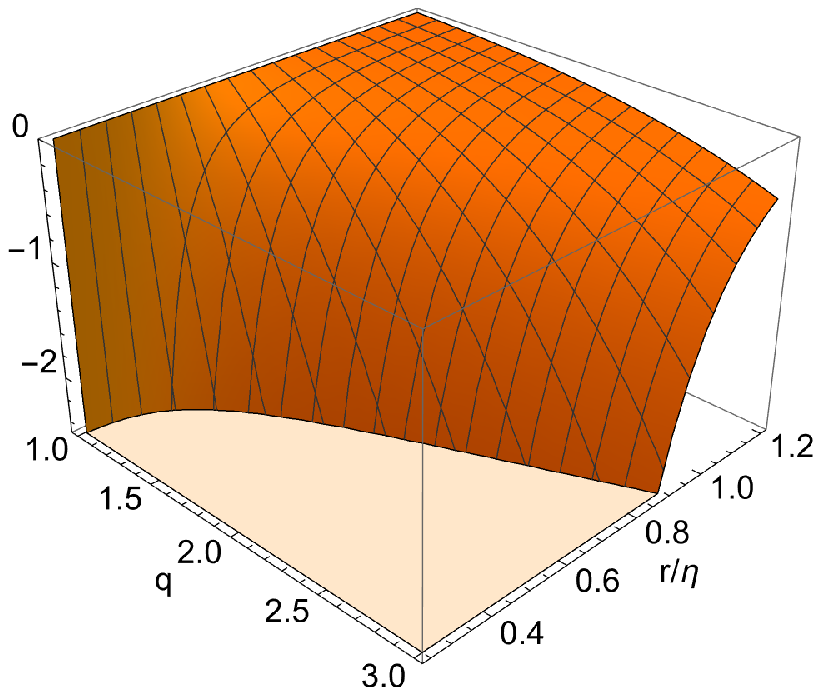,width=7.cm,height=6.cm}%
\end{tabular}%
\end{center}
\caption{The topological contributions in the vacuum energy density, $10^{2}%
\protect\alpha ^{6}\langle T_{0}^{0}\rangle _{\mathrm{t}}$ (left panel), and
in the radial stress, $10^{2}\protect\alpha ^{6}\langle T_{1}^{1}\rangle _{%
\mathrm{t}}$ (right panel), versus the proper distance from the cosmic
string (in units of the curvature radius $\protect\alpha $) and the
parameter $q$ determining the planar angle deficit. The graphs are plotted
for the model with $D=5$.}
\label{fig2}
\end{figure}

\begin{figure}[tbph]
\begin{center}
\begin{tabular}{cc}
\epsfig{figure=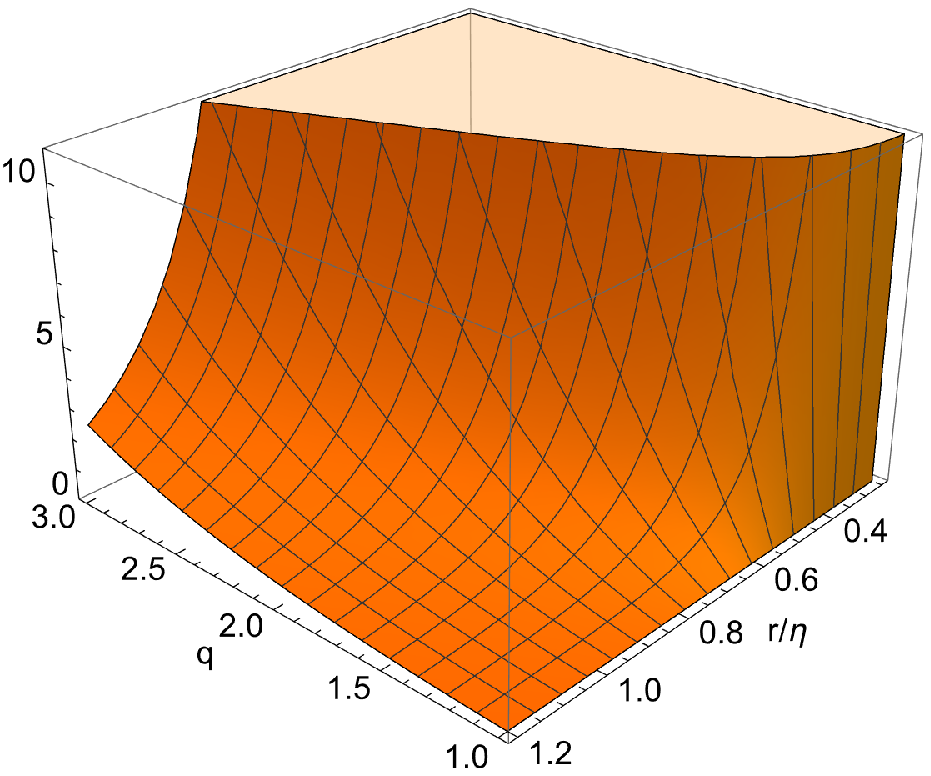,width=7.cm,height=6.cm} & \quad %
\epsfig{figure=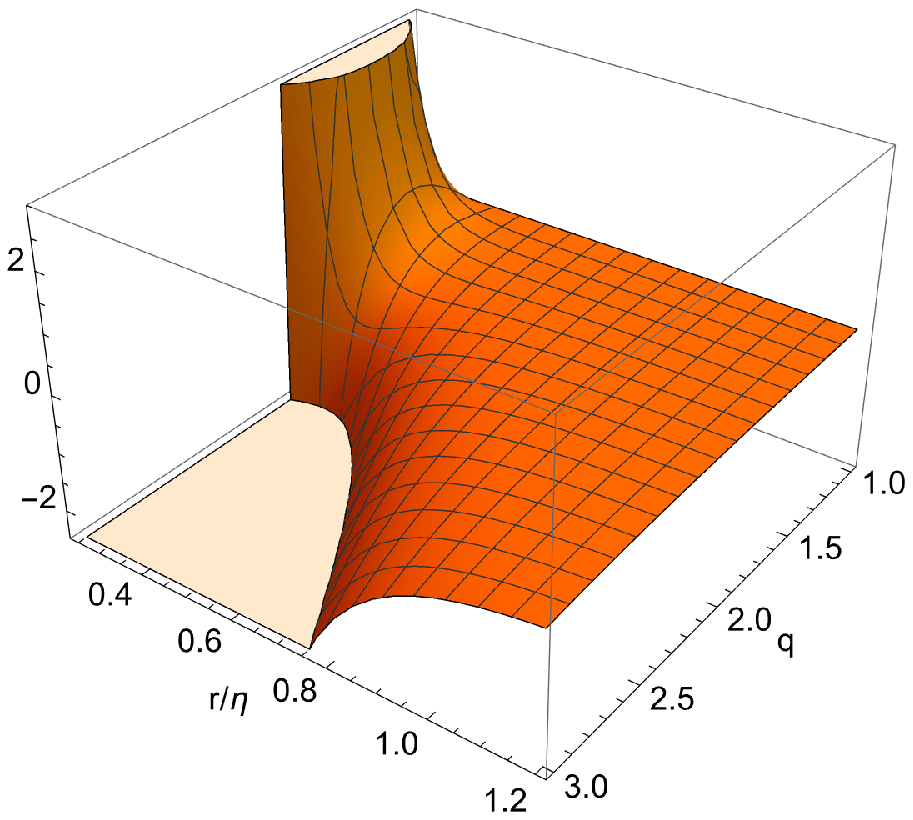,width=7.cm,height=6.cm}%
\end{tabular}%
\end{center}
\caption{The same as in Figure \protect\ref{fig2} for the azimuthal stress $%
10^{2}\protect\alpha ^{6}\langle T_{2}^{2}\rangle _{\mathrm{t}}$ (left
panel), and the axial stress $10^{2}\protect\alpha ^{6}\langle
T_{3}^{3}\rangle _{\mathrm{t}}$ (right panel).}
\label{fig3}
\end{figure}

To show the dependence on the planar angle deficit, in Figures \ref{fig2}
and \ref{fig3} we have presented the topological contributions in the
diagonal components of the vacuum energy-momentum tensor, $10^{2}\alpha
^{D+1}\langle T_{i}^{i}\rangle _{\mathrm{t}}$, as functions of the ratio $%
r/\eta $ and of the parameter $q$ for the spatial dimension $D=5$. The left
and right panels of Figure \ref{fig2} correspond to the energy density ($i=0$%
) and the radial stress ($i=1$), respectively. The left and right panels in
Figure \ref{fig3} present the azimuthal stress ($i=2$) and the axial stress (%
$i=3$). Depending on the values of $q$ and $r/\eta $, the energy density and
the axial stress corresponding to the topological contributions can be
either positive or negative. The radial azimuthal stresses are monotonic
functions of both variables. These features are in agreement with the
asymptotic analysis described above. 

\section{Energy flux}

\label{sec:Flux}

In the previous section we have investigated the diagonal components of the
energy-momentum tensor. The problem under consideration is not homogeneous
with respect to the coordinates $r$, $\eta $ and, in addition to the
diagonal components, one has a nonzero off-diagonal component $\langle
T_{0}^{1}\rangle $ that corresponds to the energy flux along the radial
direction. The respective mode sum is obtained from (\ref{Tmu}) with the
mode functions (\ref{Asig}). Again, by using (\ref{Polrel2}) for the
summation over polarizations, one gets%
\begin{equation}
\langle T_{0}^{1}\rangle =\frac{2i(D-3)q\eta ^{D+2}}{(2\pi )^{D}\alpha ^{D+1}%
}\partial _{r}\sideset{}{'}{\sum}_{m=0}^{\infty }\int d\mathbf{k}%
\int_{0}^{\infty }d\gamma \,\omega \gamma K_{\frac{D}{2}-2}(\omega \eta
e^{-\pi i/2})K_{\frac{D}{2}-1}(\omega \eta e^{\pi i/2})J_{qm}^{2}(\gamma r).
\label{T01}
\end{equation}%
The radial derivative in the right-hand side excludes the contribution of
the part corresponding to the dS geometry in the absence of the cosmic
string. That part is zero as a consequence of the problem symmetry. The
nonzero energy flux is a purely topological effect induced by the string. It
is finite for $r>0$ and the cutoff function can be removed from the
beginning.

For the further transformation of the energy flux we use the integral
representation%
\begin{eqnarray}
K_{\nu -1}(e^{-i\pi /2}\omega \eta )K_{\nu }(e^{i\pi /2}\omega \eta ) &=&-%
\frac{i}{2\eta \omega }\int_{-\infty }^{+\infty }dy\,e^{-2\nu
y}\int_{0}^{\infty }\frac{dx}{x}\left( \nu -\frac{1}{x}\right)  \notag \\
&&\times \exp \left( -x\eta ^{2}\omega ^{2}\sinh ^{2}y-\frac{1}{x}\right) .
\label{IntRepKK}
\end{eqnarray}%
The latter is obtained from the representation \cite{Wats66} 
\begin{equation}
K_{\nu }(e^{-i\pi /2}\omega \eta )K_{\nu }(e^{i\pi /2}\omega \eta ^{\prime
})=\frac{1}{2}\int_{-\infty }^{+\infty }dy\,e^{-2\nu y}\int_{0}^{\infty }%
\frac{du}{u}e^{-u/2-\omega ^{2}\beta /(2u)},  \label{IntRepKK1}
\end{equation}%
with the notation $\beta =2\eta \eta ^{\prime }\cosh (2y)-\eta ^{2}-\eta
^{\prime 2}$, taking the derivative with respect to the argument of the
second Macdonald function in (\ref{IntRepKK1}) and then passing to the limit 
$\eta ^{\prime }\rightarrow \eta $. Substituting (\ref{IntRepKK}) (with $\nu
=D/2-1$) into the expression (\ref{T01}), the integral over $\gamma $ is
evaluated by using (\ref{IntBes}). After some transformations we find%
\begin{eqnarray}
\langle T_{0}^{1}\rangle _{\mathrm{t}} &=&\frac{(3-D)q\eta }{2\left( 2\pi
\right) ^{D/2+1}\alpha ^{D+1}}\int_{0}^{\infty }du\,u^{\frac{D}{2}-1}e^{u}K_{%
\frac{D}{2}-1}(u)  \notag \\
&&\times \partial _{r}\left( 2+r\partial _{r}\right) e^{-u(r/\eta )^{2}}%
\sideset{}{'}{\sum}_{m=0}^{\infty }I_{qm}(u(r/\eta )^{2}).  \label{T012}
\end{eqnarray}%
Next, by using the formula (\ref{Sumj}), one gets%
\begin{equation}
\langle T_{0}^{1}\rangle _{\mathrm{t}}=\frac{8(D-3)r}{\left( 2\pi \right)
^{D/2+1}\alpha ^{D+1}\eta }\left[ \sum_{j=1}^{[q/2]}t^{(01)}(r/\eta ,s_{j})-%
\frac{q}{\pi }\sin (q\pi )\int_{0}^{\infty }dy\frac{t^{(01)}(r/\eta ,\cosh y)%
}{\cosh (2qy)-\cos (q\pi )}\right] ,  \label{T013}
\end{equation}%
where the notation 
\begin{equation}
t^{(01)}(x,y)=y^{2}\int_{0}^{\infty }du\,u^{\frac{D}{2}}\left(
1-uy^{2}x^{2}\right) K_{\frac{D}{2}-1}(u)e^{u-2x^{2}y^{2}u},  \label{t01xy}
\end{equation}%
has been used. Similar to the case of the diagonal components, the energy
flux density $\langle T_{0}^{1}\rangle _{\mathrm{t}}$ depends on the radial
and time coordinates through the ratio $r/\eta $ that presents to the proper
distance from the cosmic string measured in units of the dS curvature radius 
$\alpha $. An equivalent expression for the function $t^{(01)}(x,y)$ is
given by 
\begin{equation}
t^{(01)}(x,y)=\frac{y^{2}}{2}\int_{0}^{\infty }du\,u^{\frac{D}{2}%
+1}e^{u-2x^{2}y^{2}u}\left[ K_{\frac{D}{2}-2}(u)-K_{\frac{D}{2}-1}(u)\right]
.  \label{t01xyb}
\end{equation}%
This shows that this function is negative. For $\langle T_{0}^{1}\rangle _{%
\mathrm{t}}>0$ the energy flux is directed from the cosmic string and for $%
\langle T_{0}^{1}\rangle _{\mathrm{t}}<0$ it is directed towards the string.

As we could expect from the conformal relation with the problem of cosmic
string in the Minkowski bulk, the energy flux vanishes for $D=3$. For other
odd values of the spatial dimension $D$ the integral in \ (\ref{t01xy}) is
expressed in terms of the elementary functions. In particular, we get 
\begin{equation}
\langle T_{0}^{1}\rangle _{\mathrm{t}}=-\frac{c_{4}(q)}{8\pi ^{3}\alpha
^{6}\left( r/\eta \right) ^{5}},  \label{T01D5}
\end{equation}%
for $D=5$ and 
\begin{equation}
\langle T_{0}^{1}\rangle _{\mathrm{t}}=-\frac{3}{8}\frac{(r/\eta
)^{2}c_{4}(q)+c_{6}(q)}{\pi ^{4}\alpha ^{8}(r/\eta )^{7}},  \label{T01D7}
\end{equation}%
for $D=7$. In both these cases $\langle T_{0}^{1}\rangle _{\mathrm{t}}<0$.

For large values of the curvature radius, corresponding to the Minkowskian
limit, the dominant contribution in the integral representation (\ref{t01xy}%
) for the function $t^{(01)}(r/\eta ,y)$ comes from the integration range
with large values of $u$. To the leading order we get%
\begin{equation}
\langle T_{0}^{1}\rangle _{\mathrm{t}}\approx -\frac{\Gamma \left(
(D+1)/2\right) }{\left( 4\pi \right) ^{(D+1)/2}\alpha r^{D}}%
(D-3)^{2}c_{D-1}(q).  \label{T01M}
\end{equation}%
As we could expect, in the Minkowskian limit ($\alpha \rightarrow \infty $)
the energy flux vanishes. The asymptotic of the energy flux near the cosmic
string, corresponding to $r/\eta \ll 1$, is found in a similar way with the
result 
\begin{equation}
\langle T_{0}^{1}\rangle _{\mathrm{t}}\approx -\frac{(D-3)^{2}\Gamma \left(
(D+1)/2\right) }{\left( 4\pi \right) ^{(D+1)/2}\alpha ^{D+1}\left( r/\eta
\right) ^{D}}c_{D-1}(q),\;r/\eta \ll 1.  \label{T01near}
\end{equation}%
In order to find the behavior of the energy flux at large distances from the
string, $r/\eta \gg 1$, it is more convenient to use the representation (\ref%
{t01xyb}) for the function $t^{(01)}(x,y)$. In the limit under consideration
one has $x\gg 1$ and the main contribution to the integral in (\ref{t01xyb})
gives the region near the lower limit. By using the asymptotics for the
Macdonald function for small argument, we can see that, to the leading order,%
\begin{equation}
\langle T_{0}^{1}\rangle _{\mathrm{t}}\approx -\frac{(D-3)\Gamma
(D/2-1)c_{4}(q)}{8\pi ^{D/2+1}\alpha ^{D+1}\left( r/\eta \right) ^{5}}.
\label{T01far}
\end{equation}%
In the special cases $D=5,7$ this result agrees with (\ref{T01D5}) and (\ref%
{T01D7}).

The asymptotic behavior of the energy flux for large values of $q$ is
studied in the way similar to that we have used for the diagonal components.
From (\ref{t01xy}), for $xy\gg 1$ we find%
\begin{equation}
t^{(01)}(x,y)\approx \frac{\sqrt{\pi }\left( 3-D\right) }{%
2^{D/2+3}x^{D+1}y^{D-1}}\Gamma \left( \frac{D+1}{2}\right) .  \label{t01lq}
\end{equation}
In combination with (\ref{T013}) this gives the following leading term in
the expansion over $1/q$:%
\begin{equation}
\langle T_{0}^{1}\rangle _{\mathrm{t}}\approx -\frac{(D-3)^{2}\zeta \left(
D-1\right) q^{D-1}}{2^{D+1}\pi ^{3D/2-1/2}\alpha ^{D+1}\left( r/\eta \right)
^{D}}\Gamma \left( \frac{D+1}{2}\right) .  \label{T01tlq}
\end{equation}%
By taking into account that $c_{n}(q)\approx \left( q/\pi \right) ^{n}\zeta
(n)$ for $q\gg 1$, in the special cases $D=5,7$ the general estimate (\ref%
{T01tlq}) agrees with (\ref{T01D5}) and (\ref{T01D7}).

\begin{figure}[tbph]
\begin{center}
\begin{tabular}{cc}
\epsfig{figure=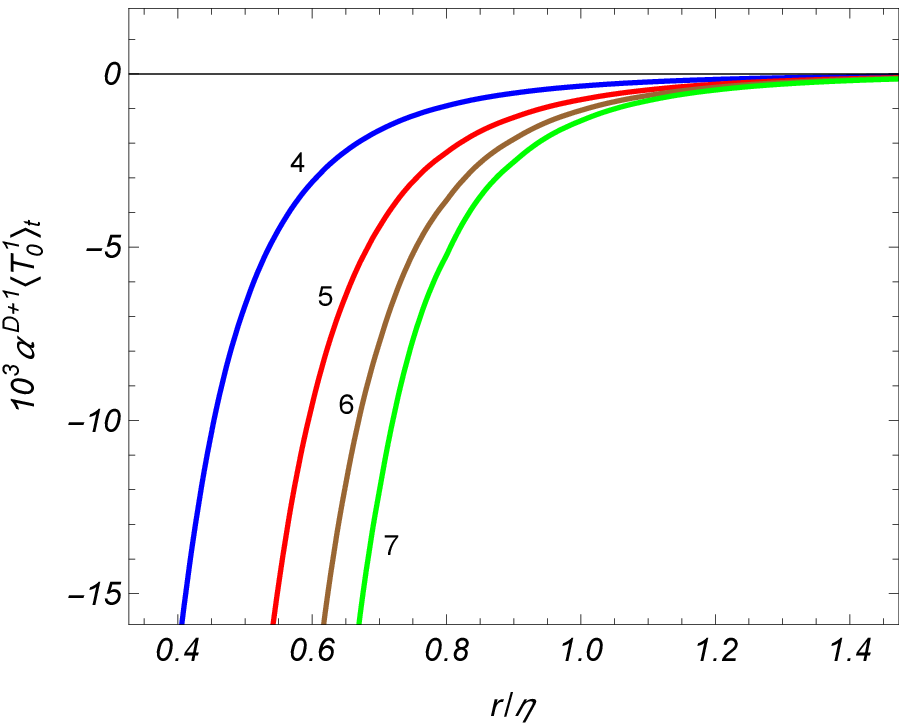,width=7.cm,height=6.cm} & \quad %
\epsfig{figure=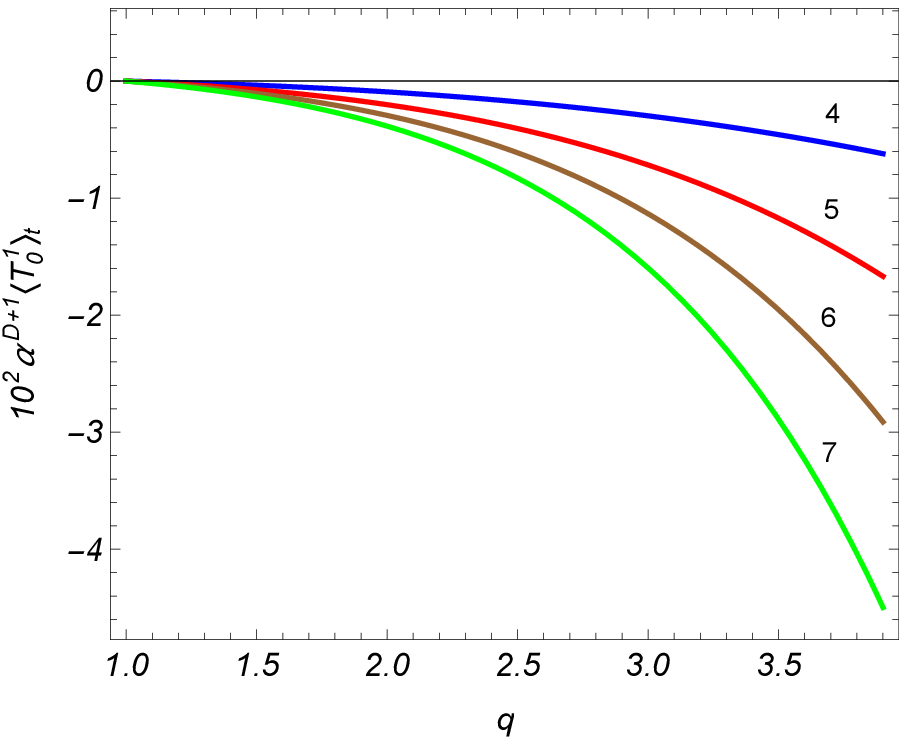,width=7.cm,height=6.cm}%
\end{tabular}%
\end{center}
\caption{The energy flux as a function of the distance from the cosmic
string (left panel) and as a function of the parameter $q$ (right panel).
The left panel is plotted for $q=1.5$ and for the right panel we have taken $%
r/\protect\eta =1$. The numbers near the curves present the values of the
spatial dimension $D$. }
\label{fig4}
\end{figure}

Figure \ref{fig4} presents the energy flux versus the radial distance (left
panel) and the parameter $q$ (right panel). The graphs on the left and right
panels are plotted for $q=1.5$ and $r/\eta =1$, respectively. The numbers
near the curves are the corresponding values of the spatial dimension. For
all the cases $\langle T_{0}^{1}\rangle _{\mathrm{t}}<0$ and the energy flux
is directed towards the cosmic string.

As an additional check for the formulas given above we can show that the
components of the topological contribution to the vacuum energy-momentum
tensor obey the covariant conservation equation $\nabla _{i}\langle
T_{l}^{i}\rangle _{\mathrm{t}}=0$. In the problem at hand two relations are
obtained between the components corresponding to the equations with $l=0$
and $l=1$. By taking into account that all the components depend on $r$ and $%
\eta $ in the form of the ratio $w=r/\eta $, they are written as%
\begin{eqnarray}
\left( w\partial _{w}+D+1\right) \langle T_{0}^{0}\rangle _{\mathrm{t}}
&=&-\left( \partial _{w}+\frac{1}{w}\right) \langle T_{0}^{1}\rangle _{%
\mathrm{t}}+\langle T_{i}^{i}\rangle _{\mathrm{t}},  \notag \\
\left( w\partial _{w}+D+1\right) \langle T_{0}^{1}\rangle _{\mathrm{t}}
&=&\left( \partial _{w}+\frac{1}{w}\right) \langle T_{1}^{1}\rangle _{%
\mathrm{t}}-\frac{1}{w}\langle T_{2}^{2}\rangle _{t}.  \label{Cons}
\end{eqnarray}%
With these relations, the components $\langle T_{i}^{i}\rangle _{t}$, $%
i=2,3,\ldots ,D$, are expressed in terms of $\langle T_{0}^{0}\rangle _{t}$, 
$\langle T_{0}^{1}\rangle _{t}$, and $\langle T_{1}^{1}\rangle _{t}$. We
recall that the components of the energy-momentum tensor were considered in
the coordinate system $(\tau ,r,\phi ,\mathbf{z})$ with the conformal time $%
\tau $. For the components in the coordinates $(t,r,\phi ,\mathbf{z})$, with 
$t$ being the synchronous time, one gets (no summation over $i$): $\langle
T_{\mathrm{(s)}i}^{i}\rangle _{\mathrm{t}}=\langle T_{i}^{i}\rangle _{%
\mathrm{t}}$ and $\langle T_{\mathrm{(s)}0}^{1}\rangle _{\mathrm{t}%
}=e^{-t/\alpha }\langle T_{0}^{1}\rangle _{\mathrm{t}}$. The topological
contribution in the vacuum energy contained in the spatial volume $V$ is
expressed as $E_{V}^{\mathrm{(t)}}=\int_{V}d^{D}x\,e^{Dt/\alpha }\langle T_{%
\mathrm{(s)}0}^{0}\rangle _{\mathrm{t}}$. Taking the region $\{r_{1}\leq
r\leq r_{2},\,0\leq \phi \leq \phi _{0},\,0\leq z^{i}\leq L_{i}\}$, $%
i=3,\ldots ,D$, as the volume $V$, from the first equation (\ref{Cons}) for
the corresponding energy $E_{V}^{\mathrm{(t)}}=E_{r_{1}\leq r\leq r_{2}}^{%
\mathrm{(t)}}$ we get%
\begin{equation}
\partial _{t}E_{r_{1}\leq r\leq r_{2}}^{\mathrm{(t)}}=\phi _{0}L_{3}\cdots
L_{D}e^{Dt/\alpha }\left[ -r\langle T_{\mathrm{(s)}0}^{1}\rangle _{\mathrm{t}%
}|_{r_{1}}^{r_{2}}+\frac{1}{\alpha }\int_{r_{1}}^{r_{2}}dr\,r%
\sum_{i=1}^{D}T_{i}^{i}\right] .  \label{Energy}
\end{equation}%
By taking into account that $\phi _{0}r_{l}L_{3}\cdots L_{D}e^{(D-1)t/\alpha
}$ is the proper surface area of the spatial hypersurface $\{r=r_{l},\,0\leq
\phi \leq \phi _{0},\,0\leq z^{i}\leq L_{i}\}$, from (\ref{Energy}) we see
that $\langle T_{0}^{1}\rangle _{\mathrm{t}}$ is the energy flux per unit
proper surface area. The second term in the square brackets of (\ref{Energy}%
) corresponds to the work done by the vacuum stresses. We recall that for
large $r$ the energy density $\langle T_{0}^{0}\rangle _{t}$ behaves as $%
1/r^{4}$ (with exception for the case $D=4$, see (\ref{T00tlarge})) and,
hence, the energy $E_{r_{1}\leq r\leq r_{2}}^{\mathrm{(t)}}$ is finite in
the limit $r_{2}\rightarrow \infty $.

In the discussion above we have considered an idealized defect with zero
thickness core. With this idealization, the topological contributions in the
diagonal components diverge on the core as $1/r^{D+1}$ and the off-diagonal
component behaves like $1/r^{D}$. Similar to the case of the Minkowskian
bulk, discussed in \cite{Beze15fc}, we can consider a simple model of finite
core of radius $a$ with the interior geometry described by the nonsingular
metric tensor $g_{il}=\left( \alpha /\tau \right) ^{2}\mathrm{diag}%
(1,-1,-u^{2}(r),-1,\ldots ,-1)$ for $r<a$. The geometry of the core is
specified by the regular function $u(r)$. The exterior metric, $r>a$, as
before, is given by (\ref{ds2}). Matching of the interior and exterior
geometries at $r=a$ is done by the Israel condition. This condition relates
the derivative of the function $u(r)$ at $r=a$ with the surface
energy-momentum tensor localized on the core boundary (see \cite{Beze15fc}
for the Minkowski bulk). For the electromagnetic modes in the region $r>a$,
instead of the Bessel function $J_{q|m|}(\gamma r)$, now the linear
combination $J_{q|m|}(\gamma r)+b_{\beta }Y_{q|m|}(\gamma r)$, with the
Neumann function $Y_{q|m|}(\gamma r)$, will appear. The coefficient $%
b_{\beta }$ is obtained from the matching conditions between the interior
and exterior modes at $r=a$. In general, it will depend on the polarization
and codifies the information about the interior geometry. The simplest model
would be the core with perfectly reflecting boundary $r=a$ that induces the
boundary condition $\left. n^{i}\,^{\ast }F_{ii_{1}\cdots
i_{D-1}}\right\vert _{r=a}=0$, where $n^{i}$ is the normal vector of
boundary and $^{\ast }F_{ii_{1}\cdots i_{D-1}}$ is the dual tensor for $%
F_{ik}$. For $D=3$ this corresponds to the boundary condition on the surface
of a perfect conductor. The effects of a conducting cylindrical shell on the
local characteristics of the electromagnetic vacuum in the geometry of a
cosmic string in (3+1)-dimensional Minkowski bulk have been studied in \cite%
{Beze07Cyl}. The influence of a cylindrical boundary in $(D+1)$-dimensional
dS spacetime, described by the line element (\ref{ds2}) with $\phi _{0}=2\pi 
$ (the cosmic string is absent) on the VEVs of the sguared electric and
magnetic fields and on the vacuum energy-momentum tensor is discussed in 
\cite{Saha16}.

In the problem at hand, the effect of a perfectly reflecting finite core of
the cosmic string on the vacuum energy-momentum tensor in the region $r>a$
can be investigated in the way similar to that used in \cite{Saha16} for the
dS bulk without a cosmic string. The coefficient $b_{\beta }$ in the
exterior mode functions is determined by the boundary condition at $r=a$.
Convenient expressions for the finite core contributions in the VEVs are
obtained by rotating the respective integrals over $\gamma $ in the complex
plane (for the corresponding procedure see \cite{Saha16}). In this way the
integrands are expressed in terms of the modified Bessel functions and the
corresponding integrals exponentially converge in the upper limit for $r>a$.
As a consequence, the vacuum energy-momentum tensor is decomposed as $%
\langle T_{i}^{l}\rangle =\langle T_{i}^{l}\rangle _{\mathrm{dS}}+\langle
T_{i}^{l}\rangle _{\mathrm{t}}+\langle T_{i}^{l}\rangle _{\mathrm{fc}}$,
where $\langle T_{i}^{l}\rangle _{\mathrm{dS}}$ is the VEV in dS spacetime, $%
\langle T_{i}^{l}\rangle _{\mathrm{t}}$ is the VEV considered in the
discussion above, and $\langle T_{i}^{l}\rangle _{\mathrm{fc}}$ is the part
induced by the finite core. When the angular deficit is absent, $\phi
_{0}=2\pi $, at large distances from the core, $r/a,r/\eta \gg 1$, the
boundary induced contribution $\langle T_{i}^{l}\rangle _{\mathrm{fc}}$ in
spatial dimensions $D>4$ decays as \cite{Saha16} $1/[\left( r/\eta \right)
^{D+2}\ln (r/a)]$ for the diagonal components and like $1/[\left( r/\eta
\right) ^{D+1}\ln (r/a)]$ for the component $\langle T_{0}^{1}\rangle _{%
\mathrm{fc}}$. For $D=4$ an additional factor $\ln (r/\eta )$ appears for
the diagonal components. We expect a similar behavior in the presence of a
cosmic string (the corresponding investigation will be presented elsewhere)
and the part $\langle T_{i}^{l}\rangle _{\mathrm{t}}$ will be dominant
compared to the contribution $\langle T_{i}^{l}\rangle _{\mathrm{fc}}$.

As it follows from the discussion above, the effects of cosmic string on the
electromagnetic vacuum in spatial dimensions $D>3$ have qualitatively new
features which are partially related to the absence of conformal invariance.
These effects may have interesting implications in string theory motivated
models where the role of cosmic strings is played by fundamental strings 
\cite{Witt85}. In those models the spatial dimension of the effective theory
depends on the specific compactification scheme and may take values in the
range $3\leq D\leq 9$. Another class of string theory inspired models with
cosmic string type structures were discussed recently within the framework
of brane inflationary models (see \cite{Cher15} for a review). The breakdown
of conformal invariance for the electromagnetic field is required in
inflationary models of the generation of large-scale magnetic fields from
quantum fluctuations in the dS expansion stage (for the discussion of
various mechanisms see, e.g., \cite{Giov04,Kand11,Durr13}). A possible
mechanism, widely considered in the literature, is based on non-minimal
interactions of the electromagnetic field with other fields. Another
possibility, related to the dynamical evolution of extra spatial dimensions
before a radiation-dominated epoch in models with $D>3$, has been discussed
in \cite{Giov00}. If the preceding stage corresponds to the inflationary
phase with dS expansion then the modifications of the electromagnetic vacuum
fluctuations discussed above will be codified in large-scale perturbations
of the magnetic field around the cosmic string in the post-inflationary
epoch. An interesting signature for the presence of extra dimensions would
be the energy flux in the radial direction (for features of inflationary
models with cosmic strings see \cite{Hind11}). Several mechanisms have been
discussed in the literature for the formation of cosmic strings during
inflation \cite{Vile00}-\cite{Cope10}. They include the direct interaction
between the inflaton and the field responsible for symmetry breaking, an
additional coupling to the background curvature and  quantum-mechanical
tunnelling \cite{Turo88,Basu91,Laza21}. Among the interesting directions for
the further research could be the back-reaction of the vacuum polarization
we have discussed here on the spacetime geometry.

\section{Conclusion}

\label{sec:Conc}

The expectation value of the energy-momentum tensor is among the central
objects in quantum field theory on curved backgrounds. In addition to being
an important local characteristic of a given state for quantum fields, it
appears as a source of the gravitational field in semiclassical Einstein
equations and determines the back reaction effects of quantum matter on the
spacetime geometry. We have discussed the combined effects of the background
geometry and topology on the VEV of the energy-momentum tensor for the
electromagnetic field in $(D+1)$-dimensional locally dS spacetime. The
nontrivial topology is generated by the presence of a topological defect
that is a generalization of a cosmic string in 4-dimensional spacetime. The
corresponding geometry is described by the line element (\ref{ds2}) where
the information on the defect is codified in the angle deficit $2\pi -\phi
_{0}$.

For evaluation of the VEV of the energy-momentum tensor we have employed the
mode-sum formula (\ref{Tmu}), with the mode functions for the vector
potential given by (\ref{Asig}). The regularization of the mode-sum is done
by introducing the cutoff function $e^{-b\omega ^{2}}$. The application of
the formula (\ref{Sumj}) allowed us to extract from the expectation values
the contributions corresponding to the pure dS bulk when the cosmic string
is absent. For spacetime points outside the defect core, $r>0$, the presence
of the cosmic string does not alter the local geometry and, hence, does not
induce new divergences in the vacuum energy-momentum tensor compared to
those for dS spacetime. Consequently, the topological contributions in the
VEVs are finite for $r>0$ and the regularization in the corresponding
expressions can be directly removed passing to the limit $b\rightarrow 0$
for the parameter in the cutoff function. The diagonal components of the
topological contribution to the VEV of the energy-momentum tensor are given
by (\ref{Tiit1}), where the function $t^{(i)}(x,y)$ is defined as (\ref{ti}%
), or equivalently, as (\ref{ti2}).\ An additional nonzero off-diagonal
component of the vacuum energy-momentum tensor, describing energy flux along
the radial direction, is given by the expression (\ref{T013}) with two
equivalent expressions, (\ref{t01xy}) and (\ref{t01xyb}), for the function $%
t^{(01)}(x,y)$ in the integrand.\ Because of the maximal symmetry of dS
spacetime, the VEVs depend on the time and radial coordinates in terms of
the proper radial distance form the cosmic string core, given by the
combination $\alpha r/\eta =re^{t/\alpha }$. For the topological
contributions we have explicitly checked the trace relation (\ref{Trace})
and the covariant conservation equation. In the problem at hand the latter
is reduced to the set of equations (\ref{Cons}).

To clarify the dependence of the topological terms on the parameters of the
problem special cases and asymptotic regions have been considered. In the
limit $\alpha \rightarrow \infty $ with fixed $t$, from the expressions for
the diagonal components of the energy-momentum tensor in dS spacetime the
cosmic string induced vacuum energy-momentum tensor is obtained in the
Minkowski bulk. In this special case the stresses along the directions
parallel to the core of the defect are equal to the energy density and they
are given by (\ref{TllM}), whereas the radial and azimuthal components are
expressed as (\ref{T11M}). All the components are monotonic functions of the
radial coordinate with power law decay $1/r^{D+1}$. The off-diagonal
component vanishes in the Minkowskian limit. The leading term in the
corresponding expansion over $1/\alpha $ is expressed as (\ref{T01M}). The
electromagnetic field is conformally invariant in 4-dimensional dS spacetime
and the corresponding topological contributions are obtained from the
expressions in the Minkowski bulk with the radial coordinate replaced by the
proper distance from the string $\alpha r/\eta $. For odd values of the
spatial dimension $D$ the expressions for the topological terms in dS bulk
are further simplified by evaluating the integrals in (\ref{ti2}) and (\ref%
{t01xyb}). The corresponding VEVs are expressed in terms of the functions (%
\ref{cn}) with even values of the index $n$. Those functions are polynomials
of degree $n$ (see (\ref{c6q}), (\ref{c8q})).

For points near the cosmic string, corresponding to small proper distances
compared to the curvature radius of dS spacetime, the influence of the
gravitational field on the diagonal components of the topological terms is
weak. The leading terms in the expansion over $r/\eta $ coincide with the
corresponding results in the Minkowski bulk where the radial distance is
replace by the proper distance $\alpha r/\eta $. The energy flux along the
radial direction is an effect induced by the gravity and the asymptotic of
the corresponding component of the energy-momentum tensor near the cosmic
string is given by (\ref{T01near}). The effects of gravity on the
topological contributions in the VEV of the energy-momentum tensor are
essential at proper distances from the cosmic string larger than the dS
curvature radius. For $r/\eta \gg 1$ and $D\geq 4$ the corresponding
asymptotic for the diagonal components is given by (\ref{Tiitlarge}). The
exception is the energy density for $D=4$ with the asymptotic behavior (\ref%
{T00tlarge}). For the energy flux density the decay at large distances is
stronger, like $1/\left( r/\eta \right) ^{5}$. Note that in contrast to the
problem of cosmic string in the Minkowski bulk, for dS background geometry
the degree of power law decay of the topological contributions at large
distances does not depend on the spatial dimension.

\section*{Acknowledgments}

A.A.S. was supported by the grant No. 21AG-1C047 of the Science Committee of
the Ministry of Education, Science, Culture and Sport RA. V.Kh.K. was
supported by the grants No. 22AA-1C002 and No. 21AG-1C069 of the Science
Committee of the Ministry of Education, Science, Culture and Sport RA.


\begin{thebibliography}{99}
\bibitem{Vile00} A. Vilenkin, E.P.S. Shellard, \textit{Cosmic Strings and
Other Topological Defects} (Cambridge University Press, Cambridge, UK, 2000).

\bibitem{Hind95} M.B. Hindmarsh, T.W.B. Kibble, Cosmic strings, Rep. Prog.
Phys. \textbf{58}, 411--562 (1995).

\bibitem{Sake07} M. Sakellariadou, Cosmic strings, Lecture Notes in Physics,
vol. 718, 247--288 (2007).

\bibitem{Cope10} E.J. Copeland, T.W.B. Kibble, Cosmic strings and
superstrings, Proc. Roy. Soc. Lond. A466, 623--657 (2010).

\bibitem{Ring10} C. Ringeval, Cosmic strings and their induced
non-Gaussianities in the cosmic microwave background, Adv. Astron. \textbf{%
2010}, 380507 (2010).

\bibitem{Lind94} A.D. Linde, \textit{Particle Physics and Inflationary
Cosmology} (Harwood Academic Publishers, Chur, Switzerland, 1990).

\bibitem{Bass07} B.A. Bassett, S. Tsujikawa, D. Wands, Inflation dynamics
and reheating, Rev. Mod. Phys. \textbf{78}, 537--589 (2007).

\bibitem{Mart14} J. Martin, C. Ringeval, V. Vennin, Encyclopedia
Inflationaris, Phys. Dark Univ. \textbf{5-6}, 75--235 (2014).

\bibitem{Ries98} A.G. Riess, et al., Observational evidence from supernovae
for an accelerating universe and a cosmological constant, Astron. J. \textbf{%
116}, 1009--1038 (1998).

\bibitem{Perl99} S. Perlmutter, et al., Measurements of omega and lambda
from 42 high-redshift supernovae, Astrophys. J. \textbf{517}, 565--586
(1999).

\bibitem{Riess07} A.G. Riess, et al., New Hubble space telescope discoveries
of Type Ia supernovae at $z \geq 1$: Narrowing constraints on the early
behavior of dark energy, Astrophys. J. \textbf{659}, 98--121 (2007).

\bibitem{Sper07} D.N. Spergel, et al., Three-year Wilkinson Microwave
Anisotropy Probe (WMAP) observations: Implications for cosmology, Astrophys.
J. Suppl. Ser. \textbf{170}, 377--408 (2007).

\bibitem{Koma09} E. Komatsu, et al., Five-year Wilkinson Microwave
Anisotropy Probe (WMAP) observations: Cosmological interpretation,
Astrophys. J. Suppl. Ser. \textbf{180}, 330--376 (2009).

\bibitem{Wein13} D.H. Weinberg, et al., Observational probes of cosmic
acceleration, Phys. Rep. \textbf{530}, 87--255 (2013).

\bibitem{Ade14} P.A.R. Ade, et al., Planck 2013 results. XVI. Cosmological
parameters, A\&A \textbf{571}, A16 (2014).

\bibitem{Giov00} M. Giovannini, Magnetogenesis and the dynamics of internal
dimensions, Phys. Rev. D \textbf{62}, 123505 (2000).

\bibitem{Giov04} M. Giovannini, The magnetized universe, Int. J. Mod. Phys.
D \textbf{13}, 391-502 (2004).

\bibitem{Atmj14} K. Atmjeet, I. Pahwa, T.R. Seshadri, K. Subramanian,
Cosmological magnetogenesis from extra-dimensional Gauss-Bonnet gravity,
Phys. Rev. D \textbf{89}, 063002 (2014).

\bibitem{Hind11} M. Hindmarsh, Signals of inflationary models with cosmic
strings, Prog. Theor. Phys. Suppl. \textbf{190}, 197--228 (2011).

\bibitem{Cope11} E.J. Copeland, L. Pogosian, T. Vachaspati, Seeking string
theory in the cosmos, Class. Quantum Grav. \textbf{28}, 204009 (2011).

\bibitem{Cher15} D.F. Chernoff, S.-H. Henry Tye, Inflation, string theory
and cosmic strings, Int. J. Mod. Phys. D \textbf{24}, 1530010 (2015).

\bibitem{Ghez02} A.M. Ghezelbash, R.B. Mann, Vortices in de Sitter
spacetimes, Phys. Lett. B \textbf{537}, 329-339 (2002).

\bibitem{Abba03} A.H. Abbassi, A.M. Abbassi, H. Razmi, Cosmological constant
influence on cosmic string spacetime, Phys. Rev. D \textbf{67}, 103504
(2003).

\bibitem{Beze03} E.R. Bezerra de Mello, Y. Brihaye, B. Hartmann, Strings in
de Sitter space, Phys. Rev. D \textbf{67}, 124008 (2003).

\bibitem{Podo04} J. Podolsk\'{y}, J.B. Griffiths, A snapping cosmic string
in a de Sitter or anti-de Sitter universe, Class. Quantum Grav. \textbf{21},
2537-2547 (2004).

\bibitem{Brih08} Y. Brihaye, B. Hartmann, Cosmic strings in a space-time
with positive cosmological constant, Phys. Lett. B \textbf{669}, 119-125
(2008).

\bibitem{Sant16} A. de P\'{a}dua Santos, E. R. Bezerra de Mello, Non-Abelian
cosmic strings in de Sitter and anti-de Sitter space, Phys. Rev. D \textbf{%
94,} 063524 (2016).

\bibitem{Davi88} P.C.W. Davies, V. Sahni, Quantum gravitational effects near
cosmic strings, Class. Quantum Grav. \textbf{5}, 1 (1988).

\bibitem{Beze09} E.R. Bezerra de Mello, A.A. Saharian, Vacuum polarization
by a cosmic string in de Sitter spacetime, J. High Energy Phys.
JHEP04(2009)046.

\bibitem{Beze10} E.R. Bezerra de Mello, A.A. Saharian, Fermionic vacuum
polarization by a cosmic string in de Sitter spacetime, J. High Energy Phys.
JHEP08(2010)038.

\bibitem{Saha17} A.A. Saharian, V.F. Manukyan, N.A. Saharyan,
Electromagnetic vacuum fluctuations around a cosmic string in de Sitter
spacetime, Eur. Phys. J. C \textbf{77}, 478 (2017).

\bibitem{Saha18Part} A.A. Saharian, V.F. Manukyan, N.A. Saharyan,
Electromagnetic vacuum densities induced by a cosmic string. Paricles 
\textbf{1}, 13 (2018).

\bibitem{Moha15} A. Mohammadi, E.R. Bezerra de Mello, A.A. Saharian, Induced
fermionic currents in de Sitter spacetime in the presence of a compactified
cosmic string, Class. Quantum Grav. \textbf{32}, 135002 (2015).

\bibitem{Brag20f} E.A.F. Bragan\c{c}a, E.R. Bezerra de Mello, A. Mohammadi,
Induced fermionic vacuum polarization in a de Sitter spacetime with a
compactified cosmic string, Phys. Rev. D \textbf{101}, 045019 (2020).

\bibitem{Brag20} E.A.F. Bragan\c{c}a, E.R. Bezerra de Mello, A. Mohammadi,
Vacuum bosonic currents induced by a compactified cosmic string in dS
background, Int. J. Mod. Phys. D \textbf{29}, 2050103 (2020).

\bibitem{Beze12AdS} E.R. Bezerra de Mello, A.A. Saharian, Vacuum
polarization induced by a cosmic string in anti-de Sitter spacetime, J.
Phys. A \textbf{45}, 115002 (2012).

\bibitem{Beze13AdS} E.R. Bezerra de Mello, E.R. Figueiredo Medeiros, A.A.
Saharian, Fermionic vacuum polarization by a cosmic string in anti-de Sitter
spacetime, Classical Quantum Gravity \textbf{30}, 175001 (2013).

\bibitem{Oliv19} W. Oliveira dos Santos, H.F. Santana Mota, E.R. Bezerra de
Mello, Induced current in high-dimensional AdS spacetime in the presence of
a cosmic string and a compactified extra dimension, Phys. Rev. D \textbf{99}%
, 045005 (2019).

\bibitem{Oliv20} W. Oliveira dos Santos, E.R. Bezerra de Mello, H. F. Mota,
Vacuum polarization in high-dimensional AdS spacetime in the presence of a
cosmic string and a compactified extra dimension, Eur. Phys. J. Plus \textbf{%
135}, 27 (2020).

\bibitem{Bell20AdS} S. Bellucci, W. Oliveira dos Santos, E.R. Bezerra de
Mello, Induced fermionic current in AdS spacetime in the presence of a
cosmic string and a compactified dimension, Eur. Phys. J. C \textbf{80}, 963
(2020).

\bibitem{Bell22AdS} S. Bellucci, W. Oliveira dos Santos, E. R. B. de Mello,
A. A. Saharian, Topological effects in fermion condensate induced by cosmic
string and compactification on AdS bulk, Symmetry \textbf{14}, 584 (2022).

\bibitem{Bell22AdSb} S. Bellucci, W. Oliveira dos Santos, E. R. Bezerra de
Mello, A. A. Saharian, Fermionic vacuum polarization around a cosmic string
in compactified AdS spacetime, J. Cosmol. Astropart. Phys. 01 (2022) 010.

\bibitem{Bell22AdSc} S. Bellucci, W. Oliveira dos Santos, E. R. Bezerra de
Mello, A. A. Saharian, Cosmic string and brane induced effects on the
fermionic vacuum in AdS spacetime, J. High Energy Phys. JHEP05(2022)021.

\bibitem{Beze22AdSt} E.R. Bezerra de Mello, W. Oliveira dos Santos, A.A.
Saharian, Finite temperature charge and current densities around a cosmic
string in AdS spacetime with compact dimension, Phys. Rev. D \textbf{106},
125009 (2022).

\bibitem{Saha16} A.A. Saharian, V.F. Manukyan, N.A. Saharyan,
Electromagnetic Casimir densities for a cylindrical shell on de Sitter
space, Int, J. Mod. Phys. A \textbf{31}, 1650183 (2016).

\bibitem{Alle86} B. Allen, T. Jacobson, Vector two-point functions in
maximally symmetric spaces, Commun. Math. Phys. \textbf{103}, 669-692 (1986).

\bibitem{Tsam07} N.C. Tsamis, R.P. Woodard, A maximally symmetric vector
propagator, J. Math. Phys. \textbf{48}, 052306 (2007).

\bibitem{Higu09} A. Higuchi, Y.C. Lee, J.R. Nicholas, More on the covariant
retarded Green's function for the electromagnetic field in de Sitter
spacetime. Phys. Rev. D \textbf{80}, 107502 (2009).

\bibitem{Yous11} A. Youssef, Infrared behavior and gauge artifacts in de
Sitter spacetime: The photon field, Phys. Rev. Lett. \textbf{107}, 021101
(2011).

\bibitem{Frob14} M.B. Fr\"{o}b, A. Higuchi, Mode-sum construction of the
two-point functions for the Stueckelberg vector fields in the Poincar\'{e}
patch of de Sitter space, J. Math. Phys. \textbf{55}, 062301 (2014).

\bibitem{Doma14} S. Domazet, T. Prokopec, A photon propagator on de Sitter
in covariant gauges, arXiv:1401.4329.

\bibitem{Nara14} G. Narain, Green's function of the vector fields on de
Sitter background, arXiv:1408.6193.

\bibitem{Glav22} D. Glavan, T. Prokopec, Photon propagator in de Sitter
space in the general covariant gauge, arXiv:2212.13982.

\bibitem{Wats66} G.N. Watson, \textit{A Treatise on the Theory of Bessel
Functions} (Cambridge University Press, Cambridge, UK, 1966).

\bibitem{Prud86} A.P. Prudnikov, Yu.A. Brychkov, O.I. Marichev, \textit{%
Integrals and Series} (Gordon and Breach, New York, 1986), Vol. II.

\bibitem{Beze10b} E.R. Bezerra de Mello, V.B. Bezerra, A.A. Saharian, V.M.
Bardeghyan, Fermionic current densities induced by magnetic flux in a
conical space with a circular boundary, Phys. Rev. D \textbf{82}, 085033
(2010).

\bibitem{Beze06} E.R. Bezerra de Mello, V.B. Bezerra, A.A. Saharian, A.S.
Tarloyan, Vacuum polarization induced by a cylindrical boundary in the
cosmic string spacetime, Phys. Rev. D \textbf{74}, 025017 (2006).

\bibitem{Frol87} V.P. Frolov, E.M. Serebriany, Vacuum polarization in the
gravitational field of a cosmic string, Phys. Rev. D \textbf{35}, 3779-3782
(1987).

\bibitem{Dowk87} J.S. Dowker, Vacuum averages for arbitrary spin around a
cosmic string, Phys. Rev. D \textbf{36}, 3742-3746 (1987).

\bibitem{Beze15fc} E.R. Bezerra de Mello, V.B. Bezerra, A.A. Saharian, H.H.
Harutyunyan, Vacuum currents induced by a magnetic flux around a cosmic
string with finite core, Phys. Rev. D \textbf{91}, 064034 (2015).

\bibitem{Beze07Cyl} E.R. Bezerra de Mello, V.B. Bezerra, A.A. Saharian,
Electromagnetic Casimir densities induced by a conducting cylindrical shell
in the cosmic string spacetime, Phys. Lett. \textbf{B645}, 245-254 (2007).

\bibitem{Witt85} E. Witten, Cosmic superstrings, Phys. Lett. \textbf{B153},
243-245 (1985)

\bibitem{Kand11} A. Kandusa, K.E. Kunze, C.G. Tsagas, Primordial
magnetogenesis, Phys. Rep. \textbf{505}, 1-58 (2011).

\bibitem{Durr13} R. Durrer, A. Neronov, Cosmological magnetic fields: their
generation, evolution and observation, Astron. Astrophys. Rev. \textbf{21},
62-109 (2013).

\bibitem{Turo88} N. Turok, String driven inflation, Phys. Rev. Lett. \textbf{%
60}, 549-552 (1988).

\bibitem{Basu91} R. Basu, A.H. Guth, A. Vilenkin, Quantum creation of
topological defects during inflation, Phys. Rev. D \textbf{44}, 340-351
(1991).

\bibitem{Laza21} G. Lazarides, R. Maji, Q. Shafi, Cosmic strings, inflation,
and gravity waves, Phys. Rev. D \textbf{104}, 095004 (2021).
\end{thebibliography}
\end{document}